\documentclass[%
 reprint,
superscriptaddress,
 amsmath,amssymb,
 aps,
prb,
]{revtex4-2}
\usepackage{placeins}
\usepackage{afterpage}
\usepackage{graphicx}
\usepackage{dcolumn}
\usepackage{bm}
\usepackage[colorlinks=true, allcolors=blue]{hyperref}

\usepackage{cleveref}
\usepackage{xcolor}

\newcommand{\review}[1]{#1}
\usepackage{physics}
\usepackage{acro}
\usepackage{caption}
\usepackage{subcaption}
\usepackage{lipsum}
\usepackage{bbm}
\usepackage{isomath}

\usepackage[centercolon]{mathtools}
\newcommand{\defeq}{\mathrel{:\mkern-0.25mu=}}
\newcommand{\eqdef}{\mathrel{=\mkern-0.25mu:}}

\DeclareMathAlphabet{\mathsfit}{T1}{\sfdefault}{\mddefault}{\sldefault}
\SetMathAlphabet{\mathsfit}{bold}{T1}{\sfdefault}{\bfdefault}{\sldefault}

\makeatletter
\newcommand{\ddV}[1]{\@ifstar{\dd \vb*{#1} \,}{\dd \vb{#1} \,}}
\makeatother

\makeatletter

\newcommand{\mat}[1]{\tensorsym{#1}}

\makeatother
\newcommand{\p}{\partial}

\renewcommand\vu[1]{\check{\vb{#1}}}

\newcommand\Zs{\mathpalette\ZZs{}}
\newcommand\ZZs[1]{\ooalign{\hss$#1\mathbf{Z}$\hss\cr\hidewidth\raisebox{0.20ex}{$#1\text{\textbf{--}}$}\hidewidth}}

\newcommand{\eg}{\epsilon}
\newcommand{\og}{\omega}

\newcommand{\sg}{\sigma}

\newcommand{\Cc}{\mathcal{C}}
\newcommand{\Dc}{\mathcal{D}}

\newcommand{\Nc}{\mathcal{N}}

\newcommand{\NN}{\mathbb{N}}
\newcommand{\ZZ}{\mathbb{Z}}
\newcommand{\RR}{\mathbb{R}}
\newcommand{\CC}{\mathbb{C}}

\newcommand{\Ai}{\mathrm{Ai}}
\newcommand{\Bi}{\mathrm{Bi}}
\newcommand{\sign}{\mathrm{sign}}

\newcommand{\at}[1]{\biggr\rvert_{#1}}

\DeclareAcronym{ASDEX}{
    short=ASDEX Upgrade,
    long=Axially Symmetric Divertor Experiment Upgrade
}

\DeclareAcronym{EBW}{
    short=EBW,
    long=electron Bernstein wave
}

\DeclareAcronym{ECRH}{
    short=ECRH,
    long=electron cyclotron resonance heating
}

\DeclareAcronym{EM}{
    short=EM,
    long=electromagnetic
}

\DeclareAcronym{AAH}{
    short=AAH,
    long=Altar-Appleton-Hartree
}

\DeclareAcronym{BC}{
    short=BC,
    long=boundary condition
}

\DeclareAcronym{CMA}{
    short=CMA,
    long=Clemmow-Mullay-Allis
}

\DeclareAcronym{GO}{
    short=GO,
    long=geometrical optics
}

\DeclareAcronym{IC}{
    short=IC,
    long=initial condition
}

\DeclareAcronym{IVP}{
    short=IVP,
    long=initial value problem
}

\DeclareAcronym{LHS}{
    short=LHS,
    long=left hand side
}

\DeclareAcronym{MGO}{
    short=MGO,
    long=metaplectic geometrical optics
}

\DeclareAcronym{NIMT}{
    short=NIMT,
    long=near identity metaplectic transform
}

\DeclareAcronym{ODE}{
    short=ODE,
    long=ordinary differential equation
}

\DeclareAcronym{ONB}{
    short=ONB,
    long=orthonormal basis
}

\DeclareAcronym{PDE}{
    short=PDE,
    long=partial differential equation
}

\DeclareAcronym{QHO}{
    short=QHO,
    long=quantum harmonic oscillator
}

\DeclareAcronym{RHS}{
    short=RHS,
    long=right hand side
}

\DeclareAcronym{RMS}{
    short=RMS,
    long=root mean square
}

\DeclareAcronym{SPA}{
    short=SPA,
    long=stationary phase approximation
}

\DeclareAcronym{SVD}{
    short=SVD,
    long=singular value decomposition,
}

\DeclareAcronym{FT}{
    short=FT,
    long=Fourier transform,
}

\DeclareAcronym{UH}{
    short=UH,
    long=upper hybrid
}

\DeclareAcronym{AAA}{
    short=AAA,
    long=adaptive Antoulas-Anderson
}

\DeclareAcronym{PIC}{
    short=PIC,
    long=particle-in-cell
}

\begin{document}

\preprint{APS/123-QED}

\title{Demonstration of Metaplectic Geometrical Optics\texorpdfstring{\\}{ }for Reduced Modeling of Plasma Waves}

\author{Rune Højlund Marholt}
\affiliation{
Section for Plasma Physics and Fusion Energy,
Department of Physics,
Technical University of Denmark,
DK-2800 Kgs. Lyngby, Denmark
}
\author{Mads Givskov Senstius}
\affiliation{
Rudolf Peierls Centre for Theoretical Physics,
University of Oxford,
Oxford OX1 3NP,
United Kingdom
}
\author{Stefan Kragh Nielsen}
\affiliation{
Section for Plasma Physics and Fusion Energy,
Department of Physics,
Technical University of Denmark,
DK-2800 Kgs. Lyngby, Denmark
}

\date{1 July 2024}

\begin{abstract}
\begin{center}
(Published in Physical Review E on 27 August 2024, Ref. \cite{Marholt2024}.\\
This is the article version before passing final review\\and without the edits from Physical Review E.)
\end{center}

\noindent
    The WKB approximation of geometrical optics is widely used in plasma physics, quantum mechanics and reduced wave modeling in general. However, it is well-known that the approximation breaks down at focal and turning points. In this work we present the first unsupervised numerical implementation of the recently developed metaplectic geometrical optics framework, which extends the applicability of geometrical optics beyond the limitations of WKB, such that the wave field remains finite at caustics. The implementation is in 1D and uses a combination of Gauss-Freud quadrature and barycentric rational function inter- and extrapolation to perform an inverse metaplectic transform numerically. The capabilities of the numerical implementation are demonstrated on Airy's and Weber's equation, which both have exact solutions to compare with. Finally, the implementation is applied to the plasma physics problem of linear conversion of X-mode to electron Bernstein waves at the upper hybrid layer and a comparison is made with results from fully kinetic particle-in-cell simulations. In all three applications we find good agreement between the exact results and the new reduced wave modeling paradigm of metaplectic geometrical optics.
\end{abstract}

\maketitle

\section{Introduction}
\noindent Ray tracing methods based on \ac{GO} are widely used for reduced wave modeling in general and in fusion plasmas in particular (Refs. \cite{Tracy2012,Batchleor1980}). Within plasma physics examples of applications include various beam based diagnostics (Refs. \cite{Donne1995,Nielsen2008}), \ac{ECRH} and current drive (Ref. \cite{Prater2008}), as well as suppression of instabilities (Ref. \cite{Nagasaki2005}). Unfortunately, the underlying eikonal (WKB) approximation behind \ac{GO} theory typically breaks down at reflection points and focal points. Such critical points are in general called caustics. Caustics are common in plasma physics and they often occur in critical regions for mode coupling, which makes them essential for various diagnostics and advanced mode coupling schemes, Refs. \cite{Preinhaelter1973,Ram2000}. Waves typically experience a natural amplification or swelling near caustics which means that several nonlinear effects can become particularly important \cite{White1974,Karney1977}. Calculating the impact of these nonlinear wave effects relies crucially on the amplitude of the wave field at the caustics, Ref. \cite{Porkolab1978}. However, the breakdown of the eikonal approximation means that the wave amplitude obtained from \ac{GO} erroneously diverges. For this reason, many wave phenomena are challenging to model properly using reduced models based on \ac{GO} and one must instead resort to computationally expensive full-wave codes \cite{Aleynikov2019,Kohn-Seemann2023}.

In the recent works of Refs. \cite{Lopez2019,Lopez2020,Lopez2021, Donelly2021,Lopez2022,Lopez2022b}, Lopez et al. proposed a new method known as \ac{MGO} for reinstating the validity of \ac{GO} in caustic regions. \ac{MGO} takes a geometrical solution strategy by recognizing that the singularities in \ac{GO} arise whenever the projection of a ray trajectory from phase space to position space is singular. This is not unlike the widely used Maslov method (Refs. \cite{Maslov1981, Keller1985}) which works by evolving the wave field in Fourier space which effectively corresponds to representing the ray trajectory in a rotated phase space, thereby eliminating some projection singularities. The \ac{MGO} method takes the Maslov method a step further by continuously applying a metaplectic transform, which corresponds to a continuous rotation of phase space in a manner such that the projection of the ray trajectory to the rotated position space is always well-defined locally. In Refs. \cite{Lopez2019,Lopez2020,Lopez2021, Donelly2021,Lopez2022,Lopez2022b}, Lopez et al. developed and demonstrated the \ac{MGO} method analytically on a few well-chosen key examples, but the method remained to be implemented in a fully automated numerical code. In this work we have developed a first of its kind automated 1D implementation of the \ac{MGO} framework \review{which differs from previous semi-analytical demonstrations of \ac{MGO} by calculating all quantities fully numerically including the inverse metaplectic transform which we shall return to in \cref{sec:MGO}}. The code is openly accessible on a repository on GitHub, Ref. \cite{GitHubRepo}. As we shall explain in \cref{sec:analytic_cont}, a particular challenge \review{in evaluating the inverse metaplectic transform} is how to analytically continue a numerical field from the real to the complex domain. This sub problem has not been addressed in the previous literature, but we suggest it can be accomplished using barycentric rational interpolation. We demonstrate the results of this newly proposed method in \cref{sec:results}.

In the following, we first present the main ideas and analytical foundations of \ac{GO} and \ac{MGO} in \cref{sec:GO,sec:MGO}. In \cref{sec:numerical_details} we then describe the numerical details of our implementation of \ac{MGO}, for which evaluating the inverse metaplectic transform is the main challenge and is achieved with barycentric rational interpolation and Gauss-Freud quadrature. In \cref{sec:airy_results,sec:weber_results} we present results of applying the method to Airy's equation and Weber's equation, which are fundamental wave physics problems with known exact analytical solutions as well as analytical \ac{MGO} solutions. Finally, in \cref{sec:XB_results} we apply the \ac{MGO} code to a caustic occurring in plasma physics when electromagnetic X-mode couples to electron Bernstein waves at the upper hybrid layer to showcase a less idealized application. The \ac{MGO} results are compared to results from fully kinetic \ac{PIC} simulations.

\section{Geometrical Optics \label{sec:GO}}
\noindent In this work we consider time-stationary scalar linear wave equations \review{of} the integrodifferential form:
\begin{gather}
\label{eq:go-wave-eq}
\int \dd {\vb{x}}'
D(\vb{x}, \vb{x}') \psi(\vb{x}') = 0,
\end{gather}
where $\vb{x}$ is the position coordinate, $D(\vb{x}, \vb{x}')$ is the wave operator kernel and $\psi(\vb{x})$ is some wave field such as a scalar electric field or a quantum mechanical wave function. To the best of our knowledge, the \ac{MGO} theory has not yet been generalized to time-dependent, vector-valued waves, but this is not an intrinsic limitation to the theory. Also note that our first iteration of a numerical implementation is only in 1D, i.e. $\vb{x} = x \in \RR$. However, we use vector notation in \cref{sec:GO,sec:MGO}, since Refs. \cite{Lopez2019,Lopez2020,Lopez2021, Donelly2021,Lopez2022,Lopez2022b} already have derived the \ac{MGO} theory in multiple dimensions.

The full integrodifferential wave equation in eq. \ref{eq:go-wave-eq} can be simplified by assuming the wave field to be \review{of} the eikonal form:
\begin{gather}
\psi(\vb{x}) = \phi(\vb{x}) e^{i \theta(\vb{x})}
\end{gather}
Here $\phi(\vb{x})$ is the envelope and $\theta(\vb{x})$ is the phase. Within the theory of \ac{GO} the envelope $\phi(\vb{x})$ is assumed to vary much more slowly compared to the phase such that high order derivatives of $\phi(\vb{x})$ can be neglected. This can also be expressed in the eikonal parameter which is assumed small:
\begin{gather}
\label{eq:eikonal-parameter}
\epsilon = \frac{1}{k L} \ll 1,
\end{gather}
where $k \sim \abs{\p_{\vb{x}} \theta}$ is the characteristic scale of variation of the phase and $1/L \sim \abs{\p_{\vb{x}} \phi}$ is the envelope variation scale. The eikonal approximation of \cref{eq:eikonal-parameter} is also known as the WKB (after Wentzel, Kramers and Brillouin) or the LG approximation (after Liouville and Green) (Ref. \cite[p. 22]{Tracy2012}) and a medium satisfying the eikonal criterion is said to be weakly inhomogenous (Ref. \cite{Mazzucato2014}). To first order in $\epsilon$ the full wave equation of \eqref{eq:go-wave-eq} can be simplified to the \ac{GO} equations (Refs. \cite{Tracy2012, Dodin2019}):
\begin{subequations}
\label{eq:go-equations}
\begin{gather}
\label{eq:go-disp-rel}
\Dc\qty[\vb{x}, \vb{k}\qty(\vb{x})] = 0. \\
\label{eq:go-envelope-eq}
\vb{v}\qty(\vb{x}) \cdot\p_{\vb{x}} \phi\qty(\vb{x})
+ \frac{1}{2} \qty[\p_{\vb{x}}\cdot \vb{v}\qty(\vb{x})] \phi\qty(\vb{x})  = 0.
\end{gather}
\end{subequations}
where $\vb{k}(\vb{x})$ and $\vb{v}(\vb{x})$ are respectively the local wave number and group velocity defined by:
\begin{gather}
\label{eq:k(x)-and-v(X)-def}
\vb{k}(\vb{x}) \defeq
\p_{\vb{x}} \theta\qty(\vb{x}), \quad
\vb{v}\qty(\vb{x}) \defeq
- \p_{\vb{k}} \Dc\qty(\vb{x}, \vb{k}) \at{\vb{k} = \vb{k}\qty(\vb{x})}.
\end{gather}
Here $\Dc(\vb{z})$ is the dispersion symbol which is a function of phase space coordinates $\vb{z}=(\vb{x},\vb{k})^T$ \review{and is assumed to be real implying that we neglect dissipation.} To be clear: $\vb{k}$ is generally any wave vector, whereas the \textit{local wave vector} $\vb{k}\qty(\vb{x})$ is a specific function of $\vb{x}$ for which \cref{eq:go-disp-rel} is satisfied. There are multiple paths for deriving the \ac{GO} equations, eq. \eqref{eq:go-equations}. A modern approach found in e.g. Refs. \cite{Dodin2019,Tracy2012} is to use Weyl symbol calculus which makes it possible to approximate the wave equation's differential operator by Taylor expanding its Weyl symbol in the eikonal parameter, eq. \eqref{eq:eikonal-parameter}. In this approach $\Dc(\vb{z})$ is found as the Wigner-Weyl transform of the wave operator (Ref. \cite{Dodin2019,Dodin2019Supplementary}). Please refer to Ref. \cite{Dodin2019} for a full derivation.

\subsection{Ray Tracing}\label{sec:RayTracing}
\noindent The \ac{GO} equations can be solved by finding phase space trajectories, $\vb{z}(\boldsymbol{\tau})=(\vb{x}(\boldsymbol{\tau}), \vb{k}(\boldsymbol{\tau}))^T$ satisfying the local dispersion relation, eq. \eqref{eq:go-disp-rel}. Such trajectories are called rays. Here $\vb*{\tau} = \qty(\tau_1, \vb*{\tau_{\perp}})^T$ where $\tau_1$ is a longitudinal time parameter and $\vb*{\tau}_\perp = \qty(\tau_2, \tau_3)^T = \qty(x_2^{(0)}, x_3^{(0)})^T$ are the perpendicular initial coordinates of the ray. Given an initial condition $\vb{z}(0,\boldsymbol{\tau}_{\perp}) = (\vb{x}_0, \vb{k}_0)^T$, a ray can be found from Hamilton's ray equations (Ref. \cite{Tracy2012}):
\begin{subequations}
\label{eq:go-ray-eqs}
\begin{gather}
\p_{\tau_1}\vb{x}(\tau_1) = - \p_{\vb{k}} \Dc(\vb{x}, \vb{k}) \\
\p_{\tau_1}\vb{k}(\tau_1) = \p_{\vb{x}} \Dc(\vb{x}, \vb{k})
\end{gather}
\end{subequations}
The dispersion symbol plays the role of the Hamiltonian. In a numerical setting in multiple dimensions, we can launch a finite family of rays on a discrete $\vb*{\tau}_{\perp}$-grid all with the same initial $x_1$-position. Thereby, we span out a region of phase space $\vb{z}(\boldsymbol{\tau})$ parameterized by $\boldsymbol{\tau} \in U, U \subseteq \RR^N$, \review{where $N$ is the number of spatial dimensions}.

The mapping $\vb{x}(\boldsymbol{\tau}) \mapsto \vb{z}(\boldsymbol{\tau}) = \qty(\vb{x}(\boldsymbol{\tau}), \vb{k}\qty[\vb{x}(\boldsymbol{\tau})])^T$ from $\vb{x}(\boldsymbol{\tau})$ to the graph of the local wave vector is called a lift. Conversely, the inverse map from $\vb{z}(\boldsymbol{\tau})$ to $\vb{x}(\boldsymbol{\tau})$ is a projection. The set of points $\qty{\vb{z}(\vb*{\tau}) \, | \, \boldsymbol{\tau} \in U}$ is an $N$-dimensional Lagrange manifold which we call the ray manifold.

\subsection{Field Construction}
\noindent Having found a ray $\vb{z}(\vb*{\tau})$ \review{satisfying \cref{eq:go-disp-rel}}, the corresponding phase and amplitude of the eikonal field \review{can be found by solving \cref{eq:go-envelope-eq,eq:k(x)-and-v(X)-def} with the explicit solutions}:
\begin{subequations}
\label{eq:go-envelope-solution}
\begin{gather}
\phi(\boldsymbol{\tau})
= \phi(0, \boldsymbol{\tau}_{\perp}) \sqrt{\frac{j(0,\boldsymbol{\tau}_{\perp})}{j(\boldsymbol{\tau})}}, \\
\qq{where}
\label{eq:go-j}
j(\boldsymbol{\tau}) \defeq \mathrm{det}[\p_{\boldsymbol{\tau}}\vb{x}(\boldsymbol{\tau})].
\end{gather}
\end{subequations}
\begin{gather}
\label{eq:go-ray-eqs-theta}
\theta\qty(\boldsymbol{\tau}) = \theta(0,\boldsymbol{\tau}_{\perp}) + \int_{0}^{\tau_1} \dd{\tau_1} \dot{\vb{x}}^T(\boldsymbol{\tau}) \vb{k}\qty(\boldsymbol{\tau})
\end{gather}
Here, det[$\cdot$] is the determinant of $\cdot$ and the dot is differentiation with respect to the first coordinate $\tau_1$. These are the essential equations of the \ac{GO} method. First, trace a set of ray phase space trajectories using eq. \eqref{eq:go-ray-eqs} to obtain the ray manifold $\qty{\qty(\vb{x}(\boldsymbol{\tau}), \vb{k}(\boldsymbol{\tau}))^T}$. Then, for each ray, determine \review{$\phi\qty(\boldsymbol{\tau})$ from eq. \eqref{eq:go-envelope-solution} and $\theta\qty(\boldsymbol{\tau})$ from eq. \eqref{eq:go-ray-eqs-theta}}. If we assume that $\vb{x}(\boldsymbol{\tau})$ has a well-defined inverse in $\vb*{\tau}(\vb{x})$, the final field as a function of position simply is:
\begin{gather}
\label{eq:go-solution-tmp}
\psi\qty(\vb{x}) = \psi\qty[\boldsymbol{\tau}\qty(\vb{x})] = \phi\qty[\boldsymbol{\tau}\qty(\vb{x})] e^{i \theta\qty[\boldsymbol{\tau}\qty(\vb{x})]},
\end{gather}

\subsection{The Caustic Problem}
\noindent For \cref{eq:go-solution-tmp} above to be meaningful, the map $\vb{x}(\boldsymbol{\tau})$ needs to be bijective such that it is invertible. This is not satisfied near turning and focal points which in both cases cause rays to cross. This caustic problem is also reflected in \cref{eq:go-envelope-solution} which diverges for $j\qty(\boldsymbol{\tau}) \to 0$. In 1D the caustic breakdown occurs if the slope $\p_{x} k(x)$ goes to infinity and therefore the local wave number function $k(x)$ does not have an explicit representation at the caustic point. In general we shall speak of the points where $j(\boldsymbol{\tau}) = 0$ as projection singularities. Within the \ac{GO} approximation the wave field diverges exactly where the projection of the ray manifold becomes singular (Ref. \cite[p. 147]{Tracy2012}).

\section{Metaplectic Geometrical Optics \label{sec:MGO}}
\noindent \ac{MGO} proposes to solve the caustic problem by continuously rotating the phase space coordinates along the ray, such that the ray manifold always has an explicit representation in the new rotated phase space coordinates. The coordinate rotations are accomplished with symplectic transforms while the corresponding transformations of the eikonal fields are accomplished with metaplectic transforms. For convenience and readability, we state in the following a few essential results on symplectic and metaplectic transforms needed to present \ac{MGO}. The reader is encouraged to consult Refs. \cite{Lopez2022,Goldstein2002,Tracy2012} for a more elaborate discussion on these topics.

\subsection{Symplectic Transforms}
\noindent Consider linear transformations of phase space coordinates \review{of} the general form:
\begin{gather}
\label{eq:mgo-sym-trfm}
\Zs =
\begin{pmatrix}
\vb{X} \\
\vb{K}
\end{pmatrix}
\defeq
\mat{S}
\,
\begin{pmatrix}
\vb{x} \\
\vb{k}
\end{pmatrix},
\end{gather}
where $\mat{S}$ is a $2N \times 2N$ matrix and $\Zs = \qty(\vb{X}, 
\vb{K})^T$ are the new phase space coordinates. Within \ac{MGO} we impose two constraints on the transformation matrix $\mat{S}$. First, the linear transformations must be canonical, i.e. they must preserve Hamilton's equations. It can be shown (see either of Refs. \cite{Goldstein2002, Tracy2012, Lopez2022}), that this is satisfied if and only if $\mat{S}$ is symplectic such that it satisfies the equation:
\begin{gather}
\label{eq:mgo-symplectic-def}
\mat{S} \mat{J}_{2N} \mat{S}^T = \mat{J}_{2N}, \qq{where}
\mat{J}_{2N} \defeq
\mqty(
\mat{0}_N & \mat{I}_N \\
- \mat{I}_N & \mat{0}_N).
\end{gather}
Here $\mat{J}_{2N}$ is known as the symplectic matrix and it is composed as a block matrix of the $N \times N$ zero matrix, $\mat{0}_N$, and the $N \times N$ identity matrix, $\mat{I}_N$.
Furthermore, we constrain ourselves to only consider rotations, why $\mat{S}$ must be orthonormal:
\begin{gather}
\label{eq:mgo-orthonormal-mat}
\mat{S}^T = \mat{S}^{-1}.
\end{gather}
The orthonormality and symplecticity requirements restricts $\mat{S}$ to be \review{of} the block form:
\begin{gather}
\label{eq:mgo-S-form}
\mat{S}
=
\begin{pmatrix}
\mat{A}     &   \mat{B} \\
-\mat{B}     &   \mat{A}
\end{pmatrix}.
\end{gather}
where $\mat{A}, \mat{B} \in \RR^{N \times N}$.

\subsubsection{Generator Formalism}
\noindent As an alternative to \cref{eq:mgo-sym-trfm}, the symplectic transformation $\vb{z} \mapsto \Zs$ can also be represented implicitly through the generator formalism. If $\mat{B}$ is invertible, the first generating function is given by (Ref. \cite[Appendix E]{Tracy2012}):
\begin{gather}
\label{eq:F1}
F_1(\vb{X}, \vb{x}) = -\frac{1}{2} \qty(\vb{X}^T \mat{A} \mat{B}^{-1} \vb{X} - 2 \vb{x}^T \mat{B}^{-1} \vb{X} + \vb{x} \mat{B}^{-1} \mat{A} \vb{x}).
\end{gather}
$F_1(\vb{X}, \vb{x})$ is defined to generate the coordinate transformation via:
\begin{gather}
\label{eq:generating-kK}
\p_{\vb{x}} F_1 = \vb{k}, \qquad
\p_{\vb{X}} F_1 = -\vb{K}.
\end{gather}

\subsection{Orthosymplectic Transformation for Singular \texorpdfstring{$\mat{B}$}{B} \label{sec:B_SVD}}
\noindent Refs. \cite{Lopez2020,Lopez2022} also treat the case where $\mat{B}$ is not invertible by considering the matrix projection of $\mat{A}$ onto the diagonalizing basis of $\mat{B}$. If $\rho$ denotes the rank of $\mat{B}$ and $\varsigma = N - \rho$ the corank, then $\mat{B}$ may be decomposed through \ac{SVD} (Ref. \cite{Nocedal2006,Lopez2022}):
\begin{gather}
\mat{B} = \mat{L}_S \tilde{\mat{B}} \mat{R}_S^T, \qq{where} \\
\label{eq:B_SVD}
\tilde{\mat{B}} =
\begin{pmatrix}
\mat{\Lambda}_{\rho \rho}   &   \mat{0}_{\rho \varsigma} \\
\mat{0}_{\varsigma \rho}   &   \mat{0}_{\varsigma \varsigma}
\end{pmatrix}, \\
\mat{L}_S^T \mat{L}_S = \mat{I}, \quad
\mat{R}_S^T \mat{R}_S = \mat{I},
\end{gather}
where the columns of $\mat{L}_S$ and $\mat{R}_S$ are the left and right singular vectors and  $\mat{L}_S$ and $\mat{R}_S$ are orthonormal. $\mat{\Lambda}_{\rho \rho}$ is a diagonal matrix. The index $\cdot_{\mu \nu}$ denotes that the matrix is of size $\mu \times \nu$. Using the requirements of orthonormality and symplecticity it is possible to show that the matrix projection of $\mat{A}$ onto the singular vectors must be block diagonal (Ref. \cite{Lopez2022}):
\begin{gather}
\label{eq:A_SVD}
\tilde{\mat{A}}
\defeq \mat{L}_S^T \mat{A} \mat{R}_S
=
\begin{pmatrix}
\mat{a}_{\rho \rho}   & \mat{0}_{\rho \varsigma} \\
\mat{0}_{\varsigma \rho}   & \mat{a}_{\varsigma \varsigma} \\
\end{pmatrix}.
\end{gather}

\subsection{Metaplectic Transforms \label{sec:MT}}
\noindent For each orthosymplectic transformation $\vb{z} \mapsto \Zs$ of phase space coordinates there exists a corresponding metaplectic transformation which defines how the field transforms from the old coordinate representation $\psi(\vb{x})$ to the new coordinate representation $\Psi(\vb{X})$. Assuming $\mat{B}$ to be invertible it is possible to show, that the representation of the metaplectic transform of $\psi(\vb{x})$ in the new $\vb{X}$-space is (Ref. \cite{Lopez2022}):
\begin{gather}
\Psi(\vb{X}) = \alpha \int \dd{\vb{x}} e^{-iF_1(\vb{X}, \vb{x})} \psi(\vb{x}),
\end{gather}
where $\alpha \in \CC$ and $F_1$ is the first generating function from \cref{eq:F1}. The metaplectic transform must be unitary and this determines $\alpha$ up to a sign. However, this leaves us with an overall sign ambiguity such that for each symplectic transformation $\mat{S}$ there are two metaplectic transformations (see Ref. \cite[p. 47-48]{Lopez2022} and Ref. \cite[p. 470]{Tracy2012} for a detailed discussion). The final integral form of the metaplectic transform of $\psi(\vb{x})$ is (Ref. \cite{Lopez2022}):
\begin{gather*}
\Psi(\vb{X}) \! = \!
\frac{
\pm e^{\frac{i}{2} \vb{X}^T \mat{A} \mat{B}^{-1} \vb{X}}
}{\scriptstyle \qty(2 \pi i)^{N/2} \sqrt{\det \mat{B}}}
\! \int \!\! \dd{\vb{x}}
\psi(\vb{x})
e^{i\qty(\frac{1}{2} \vb{x} \mat{B}^{-1} \mat{A} \vb{x} - \vb{x}^T \mat{B}^{-1} \vb{X} )}\review{.}
\end{gather*}
\review{Here $\mat{S}$ is already assumed to be of the orthosymplectic form of \cref{eq:mgo-S-form}. Under the same assumption,} the inverse metaplectic transform, which maps $\Psi(\vb{X})$ to the old representation, can be shown to be (Ref. \cite{Lopez2022}):
\begin{gather}
\label{eq:mgo-inv-MT}
\psi(\vb{x})  \! = \!
\frac{
\pm e^{-\frac{i}{2} \vb{x} \mat{B}^{-1} \mat{A} \vb{x}}
}{\scriptstyle \qty(-2 \pi i)^{N/2} \sqrt{\det \mat{B}}}
\! \int \!\! \dd{\vb{X}}
\Psi (\vb{X}) e^{i\qty(-\frac{1}{2} \vb{X}^T \mat{A} \mat{B}^{-1} \vb{X} + \vb{x}^T \mat{B}^{-1} \vb{X} )}
\end{gather}
Note how the metaplectic transform reduces to a Fourier transform in the special case where $\mat{A} = \mat{0}_N$ and $\mat{B} = \mat{I}_N$. In particular, a 1D Fourier transform is the metaplectic transform corresponding to a $90^{\circ}$ rotation in phase space.

\subsection{Geometrical Optics in Rotated Phase Space}
\review{
\noindent Importantly, Refs. \cite{Lopez2020,Lopez2022} shows that under a symplectic transformation the \ac{GO} equations, \cref{eq:go-equations}, will carry over to the new position space. To arrive at this result Refs. \cite{Lopez2020,Lopez2022} consider an eikonal form of the metaplectically transformed field in the new position space:
\begin{gather}
\label{eq:mgo-eikonal-rotated}
\Psi(\vb{X}) 
= \Phi(\vb{X}) 
\exp[i \Theta(\vb*{X}) ].
\end{gather}
Refs. \cite{Lopez2020,Lopez2022} then impose the eikonal assumption, but this time in the rotated phase space reference, i.e. $\abs{\p_{\vb{X}} \Phi} \ll \abs{\p_{\vb{X}} \Theta}$. To lowest order in the eikonal parameter the \ac{GO} equations then take the familiar form:
\begin{subequations}
\begin{gather}
\label{eq:mgo-go-eqs-rotated}
\Dc\qty[ \mat{S}^{-1} \Zs(\vb{X}) ] = 0 \\
\vb{V}(\vb{X}) \cdot \p_{\vb{X}} \Phi(\vb{X})
+ \frac{1}{2} \qty[\p_{\vb{X}} \cdot \vb{V}(\vb{X})] \Phi(\vb{X}) = 0,
\end{gather}
\end{subequations}
where $\vb{V}(\vb{X})$ is the group velocity in the new phase space coordinates. Furthermore, Ref. \cite{Lopez2020} shows that the manifold in rotated phase space is simply obtained by transforming the original representation of the manifold:
\begin{gather}
\label{eq:mgo-manifold-rotated}
\Zs\qty(\vb*{\tau}) = 
\qty(\begin{array}{cc}
\vb{X}(\vb*{\tau})\\
\vb{K}(\vb*{\tau})
\end{array})
=
\mat{S} \vb{z}(\vb*{\tau}).
\end{gather}
Therefore, in complete analogy with \cref{eq:go-ray-eqs-theta,eq:go-envelope-solution,eq:go-solution-tmp}, the new envelope and phase will have the explicit solutions in rotated phase space when away from caustics:
\begin{subequations}
\label{eq:mgo-phi-and-theta-rotated}
\begin{gather}
\Phi(\vb*{\tau})
=
\Phi(0, \vb*{\tau}_\perp) \sqrt{ \frac{
     J(0, \vb*{\tau}_\perp)
     }{
     J(\vb*{\tau})
     }
     },
\\
\text{where} \quad
J(\vb*{\tau})
=
\det[\p_{\vb*{\tau}} \vb{X}(\vb*{\tau})]. \\
\Theta\qty(\vb*{\tau}) = \Theta(0,\vb*{\tau}_{\perp}) + \int_{0}^{\tau_1} \dd{\tau_1} \dot{\vb{X}}^T(\vb*{\tau}) \vb{K}\qty(\vb*{\tau})
\end{gather}
\end{subequations}
}
\subsection{Review of \ac{MGO}}
\begin{figure*}
    \centering
    \includegraphics[width=0.65\textwidth]{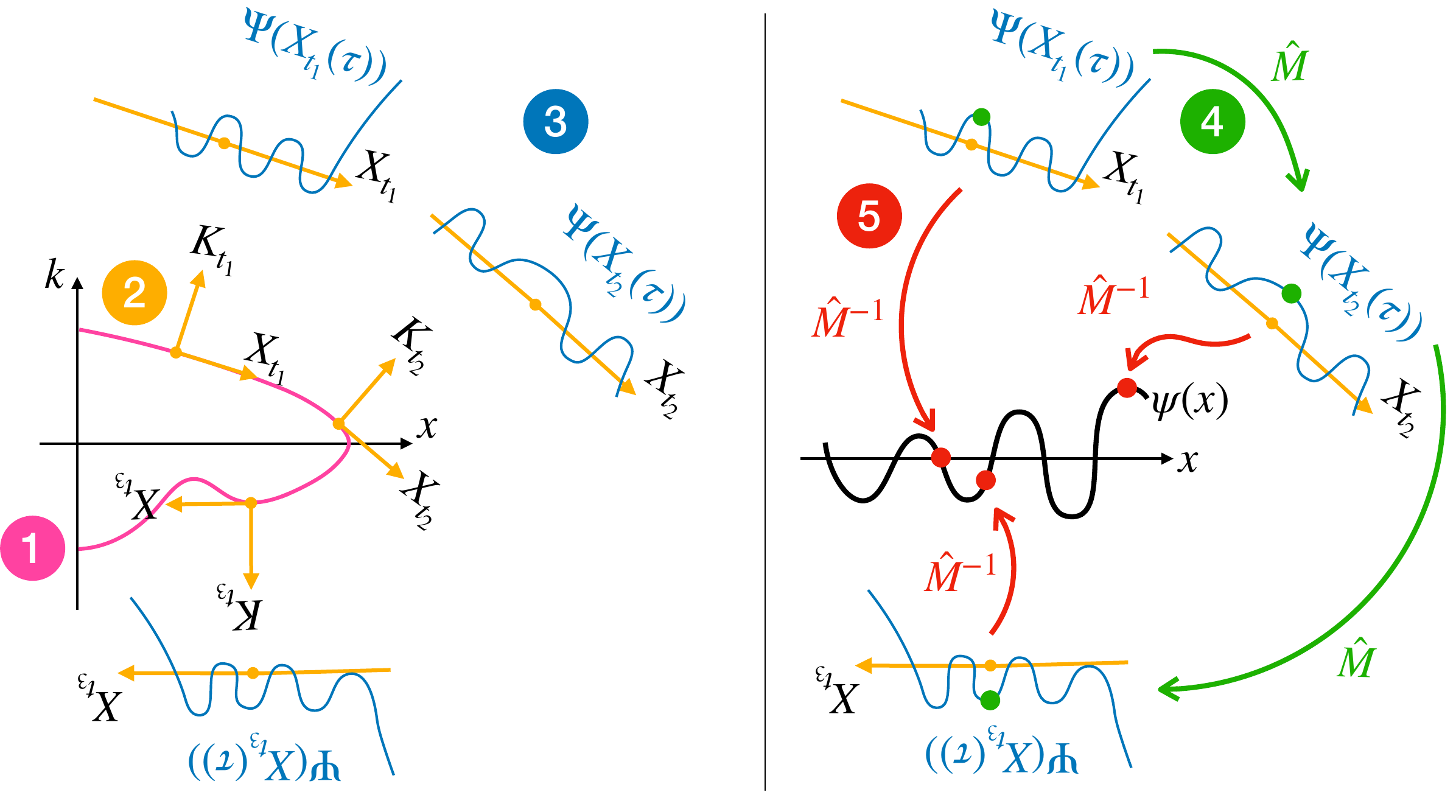}
    \caption{Conceptual illustration of the 5 steps of MGO. 1) Trace rays, 2) determine symplectic transformation $\mat{S}_t$, 3) Solve \ac{GO} in rotated phase space, 4) link the solutions together using metaplectic transforms such that the final solution is continuous 5) Inverse metaplectic transform the fields and add up the solutions from different branches.}
    \label{fig:mgo_concept}
\end{figure*}

\noindent Having presented the essentials on symplectic and metaplectic transforms, we will now present a review of the \ac{MGO} method as it was developed in Refs. \cite{Lopez2019,Lopez2020,Donelly2021,Lopez2021,Lopez2022,Lopez2022b}. This summary is in particular based on Ref. \cite{Lopez2022}, which the reader is encouraged to consult for a full derivation of the theory. The core idea of \ac{MGO} is presented in FIG. \ref{fig:mgo_concept}. The method consists of the following \review{five} steps:
\begin{enumerate}
    \item Trace a set of rays to obtain a rendering of $\vb{z}(\vb*{\tau}) \review{= \qty(\vb{x}(\vb*{\tau}), \vb{k}(\vb*{\tau}))^T}$.
    \item For each point $\boldsymbol{\tau} = \vb{t}$ on the ray manifold determine a symplectic transformation $\mat{S}_{\vb{t}}$ \review{which rotates the ray manifold representation into the new phase space coordinates:
    $$\vb{\Zs}_{\vb{t}}(\vb*{\tau}) = \qty(\vb{X}_{\vb{t}}(\vb*{\tau}), \vb{K}_{\vb{t}}(\vb*{\tau}))^T = \mat{S}_{\vb{t}} \vb{z}(\vb*{\tau}).$$
    The construction of $\mat{S}_{\vb{t}}$ is given explicitly in \cref{eq:MGO_T1,eq:MGO_Nj,eq:S_inv,eq:MGO-summary-S_t} and is designed such that the new position coordinate axes are all tangent to the ray manifold at $\boldsymbol{\tau} = \vb{t}$.}
    \item For each point $\boldsymbol{\tau} = \vb{t}$ on the ray manifold, solve the \ac{GO} phase and envelope equations \review{in the new phase space} to obtain \review{function values of $\Psi\qty(\vb{X}_{\vb{t}})$ for all $\vb{X}_{\vb{t}}(\vb*{\tau})$-points.}
    \item Ensure continuity of the solution by using a \ac{NIMT} to connect the initial conditions of the fields in rotated phase spaces together.
    \item For each point $\boldsymbol{\tau} = \vb{t}$ on the ray manifold, inverse metaplectic transform \review{$\Psi\qty(\vb{X}_{\vb{t}})$} to obtain $\psi(\vb{x}(\vb{t}))$. Sum up the field contributions from all branches of the ray manifold.
\end{enumerate}

The five steps are illustrated in FIG. \ref{fig:mgo_concept}. Note that there is a new transformation for each $\boldsymbol{\tau} = \vb{t}$ and this is reflected in the notation. For instance $\vb{X}_{\vb{t}}(\boldsymbol{\tau})$ is a function of $\boldsymbol{\tau}$ for a given fixed $\vb{t}$, while $\vb{X}_{\vb{t}}(\vb{t})$ is this function evaluated at $\boldsymbol{\tau} = \vb{t}$. Finally we denote an arbitrary coordinate set in the rotated position space as just $\vb{X}_{\vb{t}}$ and later on we shall allow $\vb{X}_{\vb{t}}$ to be complex so that we can analytically continue $\Psi(\vb{X}_{\vb{t}})$. Also note how the inverse metaplectic transform maps the function $\Psi\qty(\vb{X}_{\vb{t}})$ defined on the entire $\vb{X}_{\vb{t}}$ domain onto a single point. In the following we present a summary of the equations resulting from the five steps above.

Assume that the incoming wave field is defined on the boundary at position $x_1$:
\begin{gather}
\psi_{\text{in}}(\boldsymbol{\tau}_{\perp}) = \phi(0, \boldsymbol{\tau}_{\perp}) e^{i \theta(0,\boldsymbol{\tau}_{\perp})}.
\end{gather}
Furthermore, assume that we have an initial $\vb{z}_0$ satisfying the local dispersion relation, i.e. $\Dc(\vb{z}_0) = 0$. The first step of the procedure is to evolve the rays using \cref{eq:go-ray-eqs} to obtain $\vb{z}(\boldsymbol{\tau})$. Next, for all ray parameters $\boldsymbol{\tau} = \vb{t}$, we need to determine a symplectic rotation matrix $\mat{S}_{\vb{t}}$ such that the new position space coordinate axes span the tangent space of the ray manifold at $\vb*{\tau} = \vb{t}$. This is accomplished by defining the unit vector
\begin{gather}
\label{eq:MGO_T1}
\vu{T}_1(\vb{t}) \defeq \p_{\tau_1} \vb{z}(\vb{t}) / \norm{\p_{\tau_1} \vb{z}(\vb{t})},
\end{gather}
and then defining $\vu{T}_2(\vb{t}), \cdots, \vu{T}_N(\vb{t})$ by Gram-Schmidt orthogonalization of the Jacobian matrix $\qty[\p_{\vb*{\tau} }\vb{z}(\vb*{\tau})]^T$ (Ref. \cite{Lopez2020}). From the set of tangent vectors one can also define a symplectically dual set of normal vectors:
\begin{gather}
\label{eq:MGO_Nj}
\vu{N}_j(\vb{t}) = - \mat{J}_{2N} \vu{T}_{j}(\vb{t}).
\end{gather}
Thereby the symplectic transformation which maps $\vb{x}$ to the tangent space of the ray manifold is determined by inverting the following matrix (Ref. \cite{Lopez2020}):
\begin{gather}
\label{eq:S_inv}
\mat{S}^{-1} =
\begin{pmatrix}
\big\uparrow  &   &   \big\uparrow                      &
\big\uparrow  &   &   \big\uparrow                      \\
\vu{T}_{1}(\vb{t})  &   \cdots  &   \vu{T}_{N}(\vb{t})  &
\vu{N}_{1}(\vb{t})  &   \cdots  &   \vu{N}_{N}(\vb{t})  \\
\big\downarrow  &      &   \big\downarrow               &
\big\downarrow  &      &   \big\downarrow               \\
\end{pmatrix}.
\end{gather}
As a result, the symplectic transformation matrix is now \review{on} the orthonormal form \review{of \cref{eq:mgo-S-form}}:
\begin{gather}
\label{eq:MGO-summary-S_t}
\mat{S}_{\vb{t}} =
\begin{pmatrix}
\vb{A}_{\vb{t}}    &       \vb{B}_{\vb{t}} \\
-\vb{B}_{\vb{t}}    &       \vb{A}_{\vb{t}} \\
\end{pmatrix}.
\end{gather}
From $\mat{S}_{\vb{t}}$ we rotate the manifold representation \review{using \cref{eq:mgo-manifold-rotated}}:
\begin{gather}
\label{eq:MGO-summary-Z_t}
\Zs_{\vb{t}}(\boldsymbol{\tau}) =
\begin{pmatrix}
 \vb{X}_{\vb{t}}(\boldsymbol{\tau}) \\
 \vb{K}_{\vb{t}}(\boldsymbol{\tau})
\end{pmatrix}
\defeq
\mat{S}_{\vb{t}} \, {\mat{z}(\boldsymbol{\tau})}.
\end{gather}
By construction, this rotation ensures that the new position coordinate axes are now tangent to the rotated manifold and therefore the manifold always has an explicit representation locally. In other words, in a neighborhood of $\vb*{\tau} = \vb{t}$ the wave field is free from caustics \review{and in this neighborhood it will be justified to assume the field to be of the eikonal form of \cref{eq:mgo-eikonal-rotated}:
\begin{gather}
\Psi_{\vb{t}}\qty[\vb{X}_{\vb{t}}(\vb*{\tau})]
= \Phi_{\vb{t}} \qty[\vb{X}_{\vb{t}}(\vb*{\tau})]
\exp(i \Theta_{\vb{t}} \qty[\vb{X}_{\vb{t}}(\vb*{\tau})] ).
\end{gather}
The envelope and phase field in the rotated frame is calculated using \cref{eq:mgo-phi-and-theta-rotated}:}
\begin{subequations}
\label{eq:MGO-summary-Phi_t}
\begin{gather}
\Phi_{\vb{t}}\qty[\vb{X}_{\vb{t}}(\boldsymbol{\tau})]
=
\sqrt{ \frac{
     J_{\vb{t}}(\vb{t})
     }{
     J_{\vb{t}}(\boldsymbol{\tau})
     }
     },
\\
\label{eq:MGO-summary-J_t}
\text{where} \quad
J_{\vb{t}}(\boldsymbol{\tau})
=
\det[\p_{\boldsymbol{\tau}} \vb{X}_{\vb{t}} (\boldsymbol{\tau})].
\end{gather}
\end{subequations}
\begin{gather}
\label{eq:MGO-summary-Theta_t}
\Theta_{\vb{t}}\qty[\vb{X}_{\vb{t}}(\boldsymbol{\tau})]
=
\int_{t_1}^{\tau_1} \dd{\xi} \dot{\vb{X}}_{\vb{t}}^T(\xi, \vb*{\tau}_{\perp}) \vb{K}_{\vb{t}}(\xi, \vb*{\tau}_{\perp}).
\end{gather}
\review{Note, how the field in rotated phase space is renormalized such that
\begin{gather}
\Phi_{\vb{t}}\qty[\vb{X}_{\vb{t}}(\vb{t})] = 1,
\quad
\Theta_{\vb{t}}\qty[\vb{X}_{\vb{t}}(\vb{t})] = 0
\end{gather}
This renormalization choice is convenient, since it allows us to calculate $\Phi_{\vb{t}}$ and $\Theta_{\vb{t}}$ independently for each $\vb{t}$. To adjust for this renormalization and ensure continuity of the final wave field, a \ac{NIMT} prefactor, $\Nc_{\vb{t}}$, will be multiplied to the final wave field.}

The \review{contribution from $\Psi_{\vb{t}}\qty[\vb{X}_{\vb{t}}(\boldsymbol{\tau})]$} then needs to be brought back to the original frame. Apart from constants which we will absorb into an \ac{MGO} prefactor, $\Nc_{\boldsymbol{\tau}_i}$, the inverse metaplectic transform from \cref{eq:mgo-inv-MT} is given by:
\begin{subequations}
\label{eq:MGO-summary-Upsilon_t}
\begin{gather}
\Upsilon_{\vb{t}}
=
\int \dd{\boldsymbol{\epsilon}}
\Phi_{\vb{t}} (\boldsymbol{\epsilon})
\exp[i f_{\vb{t}}(\boldsymbol{\epsilon})] \\
\qq{where} \nonumber \\
\Phi_{\vb{t}} (\boldsymbol{\epsilon}) \defeq
\Phi_{\vb{t}}\qty[
\vb{X}_{\vb{t}}(\vb{t}) + \boldsymbol{\epsilon}
], \\
\label{eq:MGO-summary-f_t}
f_{\vb{t}}(\boldsymbol{\epsilon}) \defeq
\Theta\qty[
\vb{X}_{\vb{t}}(\vb{t}) + \boldsymbol{\epsilon}
]
-\frac{1}{2}
\boldsymbol{\epsilon}^T\vb{A}_{\vb{t}}\vb{B}_{\vb{t}}^{-1} \boldsymbol{\epsilon}
- \boldsymbol{\epsilon}^T\vb{K}_{\vb{t}}(\vb{t}),
\end{gather}
\end{subequations}
 with \review{$\vb*{\epsilon}\defeq\vb{X}_{\vb{t}}-\vb{X}_{\vb{t}}(\vb{t})$}. In section \ref{sec:evaluating-inverse-MT-integrals} we give details on how we calculate the \review{integral,} which in our current 1D implementation is only an integral over the real line. If however $\vb{B}_{\vb{t}}$ is singular, we must instead use the form:
\begin{equation}
  \begin{aligned}
\label{eq:MGO-summary-Upsilon_t_SVD}
\Upsilon_{\vb{t}}=\int_{\mathcal{C}_0} \dd{\boldsymbol{\epsilon}_\rho}
&\Psi_{\vb{t}}
\qty[
\mat{L}_{S}\left(\begin{array}{c}
\vb{X}_{\vb{t}}^\rho(\vb{t})+\boldsymbol{\epsilon}_\rho \\
\mat{a}_{\varsigma \varsigma}\cdot \vb{x}_{\varsigma}(\vb{t})
\end{array}\right)
] \\
& \times
\exp \left[-\frac{i}{2} \boldsymbol{\epsilon}_\rho^T \mat{a}_{\rho \rho} \mat{\Lambda}_{\rho \rho}^{-1} \boldsymbol{\epsilon}_\rho-i \boldsymbol{\epsilon}_\rho^T \vb{K}_{\vb{t}}^\rho(t)\right].
\end{aligned}
\end{equation}
The quantities $\mat{L}_S,\mat{a}_{\varsigma \varsigma}, \mat{\Lambda}_{\rho\rho}$ are defined through a \ac{SVD} of $\vb{B}_{\vb{t}}$ as explained in \cref{sec:B_SVD}.

The final solution to the wave field is obtained by summing up the contributions from all branches:
\begin{gather}
\label{eq:mgo-summary-psi}
\psi(\vb{x})
= \sum_{i=1}^{b} \psi(\boldsymbol{\tau}_i(\vb{x})), \qq{where}
\psi\qty[ \boldsymbol{\tau}_i(\vb{x}) ]
= \Nc_{\boldsymbol{\tau}_i} \Upsilon_{\boldsymbol{\tau}_i},
\end{gather}
where $b$ denotes the number of branches and the prefactor $\Nc_{\vb{t}}$ is given by:
\begin{subequations}
\label{eq:MGO-summary-N_t}
\begin{gather*}
\Nc_{\vb{t}}
=
\frac{
A(\vb*{\tau}_{\perp})
\exp[i \int_0^{t_1} \dd{\tau_1} \vb{k}^T(\tau_1, \vb*{\tau}_{\perp}) \dot{\vb{x}}(\tau_1, \vb*{\tau}_{\perp})]
}{
(-2 \pi i)^{\rho / 2}
e^{i \varphi/2}
\sqrt{\det \mat{\Lambda}_{\rho \rho} \det \mat{a}_{\varsigma \varsigma} \det \mat{R}_{\vb{t}}}
},
\end{gather*}
\end{subequations}
Here $\varphi(\vb{t}) \defeq \arg (\mathrm{det} (\mat{B}_{\vb{t}}))$. In our 1D implementation we require that $\varphi(t)$ must be monotonically increasing for increasing $t$. $\Nc_{\vb{t}}$ combines the prefactor from the inverse metaplectic transform and the analytic form of the \ac{NIMT} into a single analytic expression which may be evaluated independently from the integral (Ref. \cite{Lopez2022}). Note that our definition of $\Nc_{\vb{t}}$ is formulated slightly different from Ref. \cite{Lopez2022}. First, we have defined $\varphi(t)$ as the argument of $\mathrm{det} (\mat{B}_{\vb{t}})$. Since $\mathrm{det} (\mat{B}_{\vb{t}}) \in \RR$, $\varphi(t) \in \qty{n\pi \,|\, n \in \ZZ}$. By construction, $\varphi$ must be monotonically increasing as a function of $t_1$ to avoid crossing branch cuts in the square root $\sqrt{\sign \, (\mathrm{det}(\mat{B}_{\vb{t}}))} \defeq e^{i \varphi/2}$. This is directly related to the sign ambiguity of the metaplectic transform discussed in \cref{sec:MT}. Second, as opposed to Ref. \cite{Lopez2022} we define the non-zero singular values in $\mat{\Lambda}_{\rho \rho}$ to always be positive since this is a customary convention for \ac{SVD} (Ref. \cite[p. 604]{Nocedal2006}). Finally, we determine the amplitude constant $A(\vb*{\tau}_{\perp})$ by matching the final \ac{MGO} field $\psi(\vb{x})$ to the true field at some $x_1$-coordinate, where the true field is assumed to be known in the problem. Note also, that $\mat{R}_{\vb{t}}$ is the upper triangular matrix from a QR decomposition of the Jacobian matrix $\p_{\vb*{\tau}} \vb{z}(\vb*{\tau})$. This is not to be confused with the $\mat{R}_S$ from the \ac{SVD} of $\mat{B}_{\vb{t}}$.

\section{Numerical Details \label{sec:numerical_details}}
\noindent From this point onward we now restrict the position to be 1D, i.e. $N = 1$. This reflects the current state of the numerical implementation and the three examples on which we apply the code.

\subsection{Obtaining the Ray Manifold}
\noindent 
We obtained the ray manifold by integrating Hamilton's ray equations in \cref{eq:go-ray-eqs} numerically using the \ac{IVP} solver from the SciPy Library (Ref. \cite{SciPySolveIVP}) which uses a Runge-Kutta scheme. The dispersion symbol, $\Dc(\vb{z})$, depends on the particular physical system but its analytical form is known in examples below. To obtain the \ac{RHS} of \cref{eq:go-ray-eqs} we used automatic differentiation with PyTorch, Ref. \cite{PyTorch}. The \ac{IVP} solver calculated the solution at discrete points on the ray manifold, and points at equidistant ${\tau}$-values were then interpolated with a quartic polynomial. In addition to integrating the equations forward in ${\tau}$ from some initial values, we also integrate backwards by a smaller amount to get some ghost points on the manifold preceding our initial values. This makes the later step when we perform the inverse metaplectic transform numerically more robust.

\subsection{Determining Time Derivatives, the Symplectic Transformation Matrix and the Eikonal Fields}
\noindent We used a finite central difference scheme with second order precision based on the discrete known points of $\vb{z}({\tau})$ to calculate the Jacobian $j(\tau) = \p_{{\tau}} \vb{z}({\tau})$ and its' counter parts in the different rotated phase spaces $J_t(\tau) = \p_{{\tau}} \vb{\Zs_t}({\tau})$. As an alternative, one could have used the \ac{RHS} of \cref{eq:go-ray-eqs} and its' symplectically transformed analogue. However, since we used a fine $\tau$ resolution in all examples presented below, we expect the difference between these methods to be negligible compared to the larger sources of error of the \ac{MGO} method.

From $j(\tau)$ we calculated $\mat{S}_{{t}}$ at all points along the ray using \cref{eq:MGO_T1,eq:MGO_Nj,eq:S_inv}. For each point $\tau=t$, the eikonal envelope, $\Phi_t(\tau)$, was readily calculated using $J_t(\tau)$ and \cref{eq:MGO-summary-Phi_t} where we restricted the calculation to only include points on the current branch in rotated phase space. Note, we define a branch as a connected interval with constant sign of $J_t(\tau)$. For the eikonal phase, $\Theta_t(\tau)$,  in \cref{eq:MGO-summary-Theta_t}, we used numerical trapezoidal integration. With a fine $\tau$ resolution and well behaved $\dot{\vb{x}},\vb{k}$ the numerical error associated with this is expected to be negligible.  Alternatively, the phase could have been calculated by integrating the $\tau$ derivative of \cref{eq:MGO-summary-Theta_t} as an ordinary differential equation coupled to Hamilton's ray equations.

\subsection{Steepest Descent Method for the Inverse Transform \label{sec:evaluating-inverse-MT-integrals}}
\noindent
From the eikonal fields, $\Phi_t, \Theta_t$ in rotated phase space we arrive at the inverse metaplectic transform integral of \cref{eq:MGO-summary-Upsilon_t}:
\begin{gather}
\label{eq:Upsilon_t}
\Upsilon_{{t}}
=
\int_{-\infty}^\infty \dd{\epsilon}
\Phi_{{t}} (\epsilon)
e^{i f_{{t}}(\epsilon)},
\end{gather}
where \review{${\epsilon} \defeq X_{t}-X_{t}(t)$.} Due to the oscillatory term, $e^{i f_{{t}}(\epsilon)}$, the main contribution to the integral will be from the vicinity of the saddle point where \review{$\p_{{\epsilon}} f_{{t}}(\epsilon) = 0$}. Note, that by construction of $f_{{t}}(\epsilon)$ in \cref{eq:MGO-summary-f_t}, the phase factor has a saddle point exactly at $\epsilon = 0$. Attempting to evaluate the integral along the real line by simply using the trapezoidal rule will cause erroneous numerical cancellations due to the oscillatory behavior. Instead we have followed Refs. \cite{Donelly2021,Lopez2022} which propose using the method of steepest descent and Gauss-Freud quadrature. This section explains the approach.

The steepest descent method utilizes that we may deform the integral to a new contour $\gamma(l) = \epsilon \in \CC$ in the complex plane. This deformation of the integration path is allowed provided that the contributions to the integral as $|\epsilon| \to \infty$ vanishes and provided no singularities of the integrand $\Phi_{{t}}(\epsilon) e^{i f_{{t}}(\epsilon)}$ are crossed when changing the contour path (Ref. \cite[p. 158]{Gil2007}). From \cref{eq:MGO-summary-Phi_t} we see that $\Phi_{{t}}(\epsilon)$ will only have singularities along the real line (at the caustics in rotated phase space). These caustics are not crossed anew by a deformation of the contour. From the definition of $f_{{t}}(\epsilon)$ in \cref{eq:MGO-summary-f_t} we see that $f_{{t}}(\epsilon)$ is an entire function provided the eikonal phase $\Theta_{{t}}(\epsilon)$ is entire. Thus we can assume that a deformation of the integration path is in general possible. The integral is therefore:
\begin{gather}
\Upsilon_{{t}}
=
\int_{\Cc_0} \dd{\epsilon}
\Phi_{{t}} (\epsilon)
e^{i f_{{t}}(\epsilon)},
\end{gather}
The new integration contour $\Cc_0$  will be chosen as the path passing through the saddle point $\epsilon = 0$ which has the steepest descent of $\abs{e^{i f_{{t}}(\epsilon)}}$ when moving away from the saddle point. This ensures that the integral quickly converges. This is the same as finding the steepest descent of $-\Im{f_{{t}}(\eg)}$. Note, that $\Phi_{{t}}(\epsilon)$ is assumed to vary much more slowly than $e^{i f_{\vb{t}}(\epsilon)}$, so we only consider the behavior of $f_{{t}}(\epsilon)$ to be relevant in choosing the optimal integration contour.

\subsubsection{Steepest Descent Directions without Degeneracy \label{sec:steepest-descent-anal}}
\noindent A simple analysis reveals the directions of steepest descent for a non-degenerate saddle point where $f_t''(0) \neq 0$. By construction of $f_{{t}}(\epsilon)$ in \cref{eq:MGO-summary-f_t} we have:
\begin{gather}
f_{{t}}(0) = 0, \quad
f_{{t}}'(0) = 0
\end{gather}
If $f''(0) \neq 0$ we may therefore approximate $f_{{t}}(\epsilon)$ in the vicinity of $\epsilon = 0$ as:
\begin{subequations}
\begin{gather}
i f_{{t}}(\epsilon) \approx
i \frac{1}{2} f''_{{t}}(0) \epsilon^2
= \frac{1}{2} \abs{f''_{{t}}(0)} \abs{\eg}^2 e^{i \qty(\pi/2 + \alpha + 2\sigma)}, \\
\text{where} \quad
\alpha \defeq \arg f''_{{t}}(0),
\quad
\sigma \defeq \arg \epsilon.
\end{gather}
\end{subequations}
The exponential has $\cos(\pi/2 + \alpha + 2\sigma)$ as the real part. Therefore $\abs{e^{i f_{{t}}(\epsilon)}}$ is minimized when the cosine is $-1$, i.e. in the two directions where
\begin{gather}
\sigma_{\pm} = - \frac{\arg f''_{{t}}(0)}{2} - \frac{\pi}{4} \pm \frac{\pi}{2}\label{eq:theta_values}
\end{gather}
Since the cosine is $-1$ in these directions and since $f_{{t}}(0) = 0$, the real part of $f_{{t}}(\epsilon)$ will be $0$ in these directions meaning that $e^{i f_{{t}}(\epsilon)}$ will be free from oscillations if evaluated along the steepest descent direction. A similar analysis for the degenerate case where $f''_{{t}}(0) = 0, f'''_{{t}}(0) \neq 0$ gives 3 steepest descent directions and so forth for higher degeneracy orders.

\subsubsection{Gauss-Freud Quadrature}
\noindent Now, assume we have found a parameterization $\gamma : \RR \to \CC$ of the contour such that the contour integral is:
\begin{gather}
\Upsilon_{{t}}
=
\int_{-\infty}^{\infty} \dd{l}
\gamma'(l)
\Phi_{{t}} \qty[\gamma(l)]
e^{i f_{{t}}\qty[\gamma(l)]}
\end{gather}
Further, assume that the saddle point is reached at $l=0$, i.e. $\gamma(0) = 0$. If the saddle point is degenerate, $\Cc_0$ may have a kink at the saddle point and it is therefore convenient to split $\gamma$ into two functions:
\begin{gather}
\gamma(l) \defeq
\begin{cases}
\gamma_{-}(l)   \quad   l \leq 0 \\
\gamma_{+}(l)   \quad   l > 0
\end{cases}.
\end{gather}
The contour integral can then be written as a single integral from $0$ to $\infty$:
\begin{align}
\Upsilon_{{t}}
=
\int_{0}^{\infty} \dd{l}
\bigg(
& \gamma'_{-}(-l)
\Phi_{{t}} \qty[\gamma_{-}(-l)]
e^{i f_{{t}}\qty[\gamma_{-}(-l)]}
\\
& +
\gamma'_{+}(l)
\Phi_{{t}} \qty[\gamma_{+}(l)]
e^{i f_{{t}}\qty[\gamma_{+}(l)]}
\bigg)
\end{align}
Using Gauss-Freud quadrature, an integral \review{of} the form above can be approximated as a finite sum:
\begin{gather}
\int_{0}^{\infty} \dd{l} h(l) \approx \sum_{j=1}^{n} w_j \frac{h(l_j)}{\omega(l_j)},
\quad
\text{where}\\
\omega(l) = e^{-l^2}.
\end{gather}
Please refer to Refs. \cite{Donelly2021,Gil2007} for more details on the Gauss-Freud quadrature method. To reduce the error in the quadrature approximation, we need $h(l)/\omega(l)$ to be well-approximated by a $2n-1$ degree polynomial (Ref. \cite{Gil2007}). We shall therefore introduce a constant scaling, $\lambda$, of the parameterization, such that the integral becomes:
\begin{gather}
\Upsilon_{{t}}
=
\int_{0}^{\infty} \dd{l} h(l),
\qq{where} h(l) \defeq h_{-}(l) + h_{+}(l), \\
h_{\pm}(l) =
\lambda
\qty(
\gamma'_{\pm}(\pm\lambda l)
\Phi_{{t}} \qty[\gamma_{\pm}(\pm\lambda l)]
e^{i f_{{t}}\qty[\gamma_{\pm}(\pm \lambda l)]}
).
\end{gather}
\review{To choose $\lambda$ we note that $f_t\qty[\gamma(l)]$ is purely imaginary and increasing and then Taylor expand the imaginary part around $l=0$:
\begin{subequations}
\begin{gather}
f_t\qty[\gamma(l)]
= i\qty[ \qty(\frac{l}{\lambda_2})^2 + \qty(\frac{l}{\lambda_3})^3 + \cdots] \\
\text{where} \qquad
\lambda_m \defeq \abs{\frac{1}{m!}\p_{l}^m \Im{f_t\qty[\gamma(l)]}\at{l=0}}^{-1/m}
\end{gather}
\end{subequations}
The absolute value is included to stress that we assume the $m$'th derivative of $\Im f_t \qty[\gamma(l)]$ to be positive. We propose defining the global length scale as $\lambda = \lambda_m$, where $m$ is the lowest positive integer for which $\lambda_m \leq \lambda_{m+1}$. With this definition, we will approximately have (in the vicinity of the saddle point):
\begin{gather}
f_t\qty[\gamma(\lambda l)] \approx il^m
\end{gather}
And thereby
\begin{gather}
\frac{h_{\pm}(l)}{\og(l)}
\approx
\lambda \gamma_{\pm}'(\pm \lambda l)
\Phi_t \qty[\gamma_{\pm} \qty(\pm \lambda l)] e^{l^2 - l^{m}}
\end{gather}
The Gauss Freud quadrature is therefore appropriate if $\Phi_t(l)e^{l^2 - l^{m}}$ is well-approximated by a $(2n-1)$ degree polynomial, where $m$ is the order of degeneracy. This should work especially well for non-degenerate saddle points, where $m=2$.}

\subsubsection{Steepest Descent Angle Update Algorithm}
\noindent \Cref{eq:theta_values} is only accurate far from caustics, since the saddle point becomes degenerate at the caustic. Instead of using \cref{eq:theta_values} we have therefore implemented an algorithm similar to, but slightly different from, the angle update algorithm of Ref. \cite{Donelly2021}. First, we assume that the directions of the contours are unchanged as we move away from the saddle point such that
\begin{gather}
\gamma(l) \defeq
\begin{cases}
|l|e^{i\sigma_-}   \quad   l \leq 0 \\
|l|e^{i\sigma_+}   \quad   l > 0
\end{cases}.
\end{gather}
We start an iteration on each branch of the manifold at the $\tau$-value with maximal value of $\abs{j(\tau)}$. This is to ensure that we are as far from the caustics as possible such that we can use \cref{eq:theta_values} as our initial values of $\sigma_{\pm}$. Then, for each point on the ray manifold we minimize $-\mathrm{Im}\{f(\epsilon)\}$ on a circle of radius $L$ and select the closest minimal loci $\sigma_{\pm}$ relative to the $\sigma_{\pm}$-values found at the previous $\tau$ step. In doing so, we avoid $\sigma_+=\sigma_-$ which would not be a valid deformation of the contour. For the radius of the circle, we choose $L=l_1\lambda$ where $l_1$ is the lowest order node in the Gauss-Freud quadrature.

\subsection{Analytic Continuation to the Complex Plane \label{sec:analytic_cont}}
\noindent To evaluate the integrand along the steepest descent contour we need to know the values of the integrand in the complex domain. This is no problem in an analytic implementation of \ac{MGO}, but in a numerical treatment we only know the function values of $\Psi_{{t}}\qty[\epsilon + {X}_{{t}}({t})]$ on the real domain after having calculated the phase and envelope using \cref{eq:MGO-summary-Phi_t,eq:MGO-summary-Theta_t}. To solve this problem, we use barycentric rational interpolation \review{(see Refs. \cite{Berrut2014, Nakatsukasa2018, Hofreither2021, Trefethen2023})} of the numerical signal of $\Psi_{{t}}, f_{{t}}$. In a barycentric rational interpolation, a function $f(z)$ of a complex variable $z$ is represented as the ratio of two partial fractions:
\begin{gather}
r(z)=\frac{n(z)}{d(z)}=\sum_{j=1}^m \frac{w_j f_j}{z-z_j} 
\bigg/ \sum_{j=1}^m \frac{w_j}{z-z_j},
\end{gather}
where $f_j=f(z_j)$ are known sampled values of the function and $w_j$ are weights which must be chosen. In this form, one can see that $r\to f_j$ for $z\to z_j$ why defining $r(z_j) \defeq f_j$ for all $j$ is meaningful such that $r(z)$ is continuous and takes the sampled values at all interpolation points. Multiplying the nominator and denominator with the node polynomial $\ell$:
\begin{gather}
\ell(z) = \prod_{j=1}^m \qty(z - z_j),
\end{gather}
shows that the barycentric representation is in fact a rational function, i.e. a quotient of two polynomials, Ref. \cite{Nakatsukasa2018}. The barycentric form is less prone to numerical overflow compared to the raw rational form, Ref. \cite{Berrut2014}. A well known special case of barycentric interpolation is Lagrange polynomial interpolation, which uses $w_j=\prod_{k\neq j}(z_j-z_k)^{-1}$. Rather than this choice of weights, we use the recently proposed \ac{AAA} algorithm to find the best interpolation. \review{The algorithm is described in Ref. \cite{Nakatsukasa2018} and the Python implementation which we use is due to Refs. \cite{Hofreither2021, baryratGitHub}.} Although the algorithm generally performs well when interpolating, the precision when extrapolating an oscillatory function is generally acceptable one wavelength away from the interpolation domain, Ref. \cite{Trefethen2023}. An issue in using the \ac{AAA} algorithm for extrapolation is however that we only know the function values in a limited interval $\epsilon \in [a,b]$ along the real line. Thus we are not guaranteed to have data in a sufficiently large radius around $\epsilon=0$ for an interpolation to be valid. Furthermore, there may also be caustics at $\epsilon\neq0$ in the rotated frame which can cause numerical challenges. With these considerations, we pick a maximal value $M$ beyond which we do not extrapolate. The value is chosen based on the wave number in the rotated phase space such that
\begin{gather}
M=\epsilon_{\mathrm{max}} + \frac{\pi}{K_t(t)},
\end{gather}
where $\epsilon_{\mathrm{max}}\eqdef \mathrm{min}(|a|, |b|)$. The maximal number of quadrature nodes we use is 10, but we choose the highest number of nodes, $n$, such that $\lambda l_n<M$, where $l_n$ is the position of the $n$'th quadrature node. In the worst case where even the first quadrature node is outside the extrapolation domain, i.e. $\lambda l_1<M$, we instead change the scaling $\lambda$ to make the nodes fit inside the extrapolation domain. There is a trade-off here; more nodes improves the accuracy of the integral, but at the same time requires the barycentric rational interpolation to be valid in a larger radius in the complex domain. The choices above are an attempt to balance these considerations in an automated fashion.

\subsection{Constructing the field}
\noindent Once the inverse metaplectic transform has been applied, the branches of different sign of $j(\tau)$ must be superimposed in order to produce the total field. As different branches may be known at different positions and with varying resolution, we here apply interpolation once again to evaluate the different branches in the same points. For this purpose we use a linear interpolation scheme on the different branches of the wave field in the original frame.

\section{Results \label{sec:results}}
\noindent Having outlined the theory of \ac{MGO} and the numerical details of the implementation we will now proceed to showcase \ac{MGO} in action. The only type of caustic occurring in 1D is the fold caustic and we shall see three examples of this. First we examine the Airy and Weber equation. Both problems are fundamental examples of caustics and have been treated analytically in the previous work by Ref. \cite{Lopez2020}. Lastly, we demonstrate the numerical implementation on a fold caustic in connection with X-B mode coupling in a hot magnetized plasma. This problem has not previously been treated using \ac{MGO} so we compare the results with \ac{PIC} simulations.

\subsection{Airy's Equation \label{sec:airy_results}}
\begin{figure*}[htb]
    \centering
    \includegraphics[width=\textwidth]{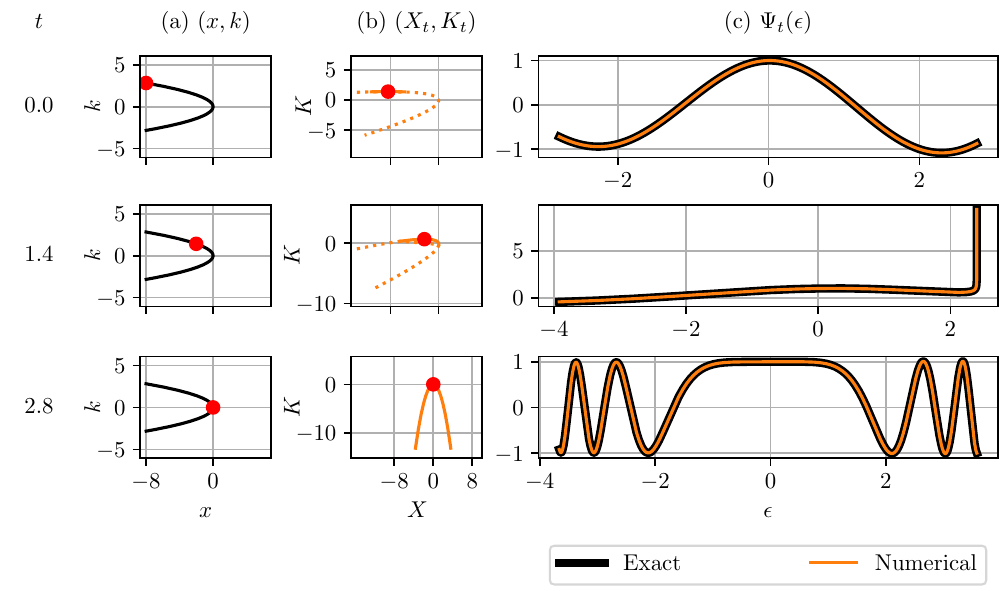}
    \caption{Illustration of the \ac{MGO} algorithm at different time points for the Airy equation example. (a) Ray manifold with red dot indicating phase space location at time $t$. (b) Current branch of symplectically transformed ray manifold. The solid orange line is the part of the current branch actually used for the Gauss-Freud quadrature integration. (c) Metaplectically transformed eikonal wave field $\Psi_t(\epsilon) = \Phi_t(\epsilon) e^{i \theta_t(\epsilon)}$ as a function of $\epsilon = X_t(\tau) - X_t(t)$ corresponding to symplectic transformation at time $t$. For (c) we have shown both the exact closed-form \ac{MGO} result from \eqref{eq:Airys_field} and the result from our numerical implementation. The core ray tracing started at $x(0)=-8$ and was automatically stopped when returning to its' starting position. Besides this a ghost point tracing was carried out to enable the calculation of the eikonal fields at the boundaries.}
    \label{fig:airy_steps_1}
\end{figure*}
\begin{figure*}[htb]
    \centering
    \includegraphics[width=\textwidth]{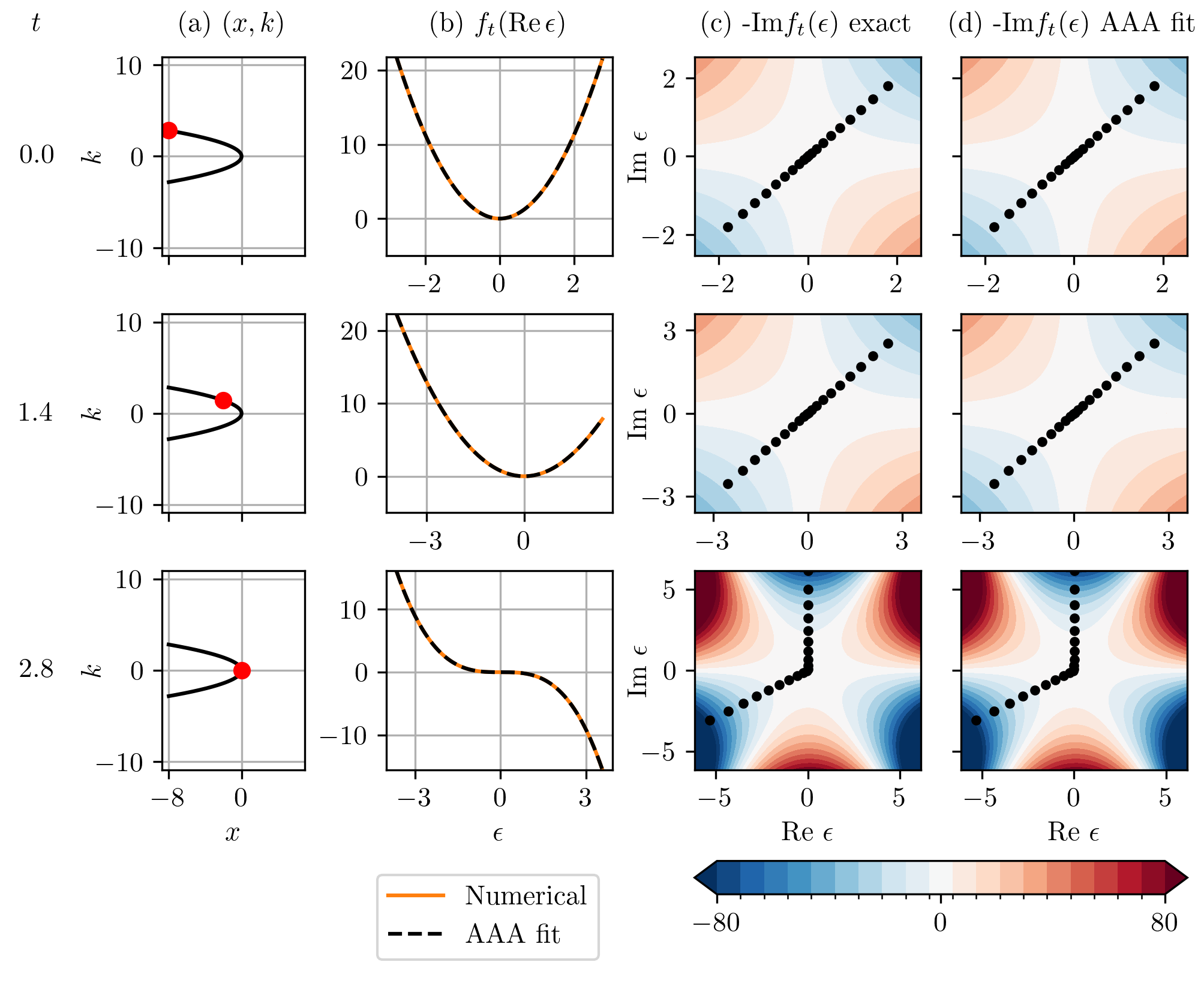}
    \caption{Inspection of barycentric rational interpolation needed for the analytic continuation in order to evaluate the steepest descent integral as part of the inverse metaplectic transform step for the Airy equation example. (a) Ray manifold with red dot indicating phase space location at time $t$, (b) Metaplectically transformed phase factor, $f_t$, as function of $\epsilon = X_t - X_t(t) \in \RR$, (c) exact negative imaginary part of the analytically continued phase factor from Eq. \eqref{eq:Airys_ft} evaluated on $\epsilon \in \CC$, (d) "-Im" part of barycentric rational interpolation of $f_t$ evaluated on $\epsilon \in \CC$. In (b) we have included the numerical result (which in FIG. \ref{fig:airy_steps_1} was found to agree with the exact result along Re $\epsilon$.), and the barycentric rational interpolation. For (c) and (d) we have also shown the Gauss quadrature node loci ($\epsilon_j = \lambda l_j e^{\sg_\pm}$) along the steepest descent contour used to evaluate the integral of the inverse metaplectic transform.}
    \label{fig:airy_steps_2}
\end{figure*}
\noindent As a first example consider Airy's Equation
\begin{gather}
\label{eq:Ai_eq}
\p_x^2 \psi(x) - x \psi(x) = 0.
\end{gather}
Insertion of a plane wave ansatz or alternatively taking the Wigner-Weyl transform of the wave operator gives us the local dispersion relation for Airy's equation:
\begin{gather}
\label{eq:example-airy-D}
\Dc(x, k) = - k^2 - x = 0
\end{gather}
A longer analytical analysis of the problem using the formulas from \cref{sec:MGO} gives the following fields needed for the contour integrals (Refs. \cite{Lopez2020,Lopez2022})
\begin{gather}
    f_t(\epsilon)=\Theta[X_t(t)+\epsilon]+k(t)\epsilon^2-\dfrac{k^2(t)}{\vartheta_t}\epsilon,\label{eq:Airys_ft}\\
    \Theta_t[X_t(t)+\epsilon]=\dfrac{8k^4(t)-\vartheta_t^4}{8k^2(t)\vartheta_t}\vartheta_t + \dfrac{1}{4k(t)}\epsilon^2 \notag\\
    +\; \dfrac{\vartheta_t^6-[\vartheta_t^4-8k(t)\vartheta_t\epsilon]^{3/2}}{96k^3(t)},
\end{gather}
where 
\begin{gather*}
\vartheta_t=\sqrt{1+4k^2(t)},\\
k(t)=\sqrt{-x(t)}, \quad
x(t)=-(\sqrt{x_0}-t)^2
\end{gather*}
and $x_0$ is the initial position of the traced rays. Finally, the field in the rotated frame is given by
\begin{gather}
    \Psi_t[X_t(t)+\epsilon]=\dfrac{\vartheta_t}{\sqrt{4k(t)k(\tau)+1}}e^{i\Theta_t[X_t(t)+\epsilon]}.\label{eq:Airys_field}
\end{gather}
These analytical results will not be used to generate the numerical results but are used for bench marking against in FIG. \ref{fig:airy_steps_2}. Airy's equation has a fold caustic at $x=0$ where the wave number $k$ must vanish for the dispersion relation to be satisfied. In quantum mechanics such fold caustics are encountered at turning points between the classical and non-classical region of a potential barrier (Ref. \cite[Ch. 9]{Griffiths2018}). In plasma physics, Airy's equation for instance arise when an O-mode meets its' cutoff at the critical density $n_c = \epsilon_0 m_e \omega^2 / (e^{2})$. In a 1D plasma physics context, the wave field $\psi(x)$ is therefore the electric field $E(x)$. Since \cref{eq:Ai_eq} is a second-order differential equation it has two linearly independent basis solutions: $\Ai(x), \Bi(x)$. Both are special cases of the modified Bessel functions which are part of the larger family of hypergeometric functions, Ref. \cite{Gil2007}. The solution to \cref{eq:Ai_eq} is a linear combination of these two, but since $\Bi(x)$ diverges for $x \to \infty$ the solution we are interested in is the Airy function of the first kind. Up to a normalization constant, the solution therefore is:
\begin{gather}
\psi(x) = \Ai(x).
\end{gather}
For the numerical solutions, we first trace a ray starting at $x=-8$, which propagates to $x=0$ where it is reflected and returns to $x=-8$, at which point we automatically terminate the tracing. Using the formulas in section \ref{sec:RayTracing} along with the initial condition $\psi(-8)=\Ai(-8)$, we obtain the \ac{GO} solution. For the \ac{MGO} solution, we trace backwards to $x \approx -13$ to provide ghost points for the calculations. Except for this fact, we use the same ray trajectory as for \ac{GO} case.

Intermediate results from the \ac{MGO} algorithm for a few selected time points are shown in FIG. \ref{fig:airy_steps_1}. The first time point is at the beginning, far from the caustic, the second is near the caustic and the final time point is at the caustic. From FIG. \ref{fig:airy_steps_1} (b) it is clearly visible how at all times the rotated manifold is always tangent to the $X$-axis at $\tau=t$.  As a result the metaplectically transformed fields plotted in FIG. \ref{fig:airy_steps_1} (c) are all free from caustics in a neighborhood of $\epsilon=0$. However, singularities may still appear further away, as is the case for $t=1.4$. At $t=2.8$ we meet the caustic of the original frame and the ray manifold is rotated $90^\circ$. Therefore, the metaplectically transformed eikonal field shown at the lowest plot of FIG. \ref{fig:airy_steps_1} (c) is actually a Fourier transform of the Airy function. Importantly we see an excellent agreement between the exact analytical fields from \cref{eq:Airys_field} and our numerical fields on FIG. \ref{fig:airy_steps_1} (c).

On FIG. \ref{fig:airy_steps_2} we evaluate the analytic continuation step for the phase factor $f_t(\epsilon)$ of \cref{eq:MGO-summary-f_t} needed in the inverse metaplectic transform. We focus in  FIG. \ref{fig:airy_steps_2} on the same three time points as in FIG. \ref{fig:airy_steps_1}. At all times, the barycentric rational interpolation using the AAA algorithm appears to agree very well with $f_t$ along the real line. To investigate the analytical continuation we show the negative imaginary part of $f_t(\epsilon)$ in FIG. \ref{fig:airy_steps_2} (c)-(d) in the complex plane. We see no visible error between the exact result and the AAA fit in the complex plane either. At $t=2.8$ where we meet the caustic in the original frame, we see how the degeneracy of the saddle point gives three possible steepest descent directions as explained in section \ref{sec:steepest-descent-anal}. At all time points the algorithm for finding the directions of the steepest decent contours along with the straight line assumption for the contours appear to place the Gauss-Freud quadrature nodes close to the intended contour.

As the final steps of the \ac{MGO} algorithm we calculate the prefactor stemming from the \ac{NIMT} using the initial condition. Finally the branch contributions are interpolated and superimposed to give the final result shown in FIG. \ref{fig:airy}. In FIG. \ref{fig:airy} we compare the three solutions along the original position axis. Of course, neither of the \ac{GO} and \ac{MGO} fields extend into the evanescent region $x>0$. Far from the caustic for $x<0$, all solutions agree very well. Near the caustic, the \ac{GO} solution diverges as anticipated while the \ac{MGO} solution stays close to the exact solution everywhere.

\begin{figure}[htb]
\centering
\begin{subfigure}[t]{\linewidth}
    \caption{}
    \includegraphics[width=\linewidth]{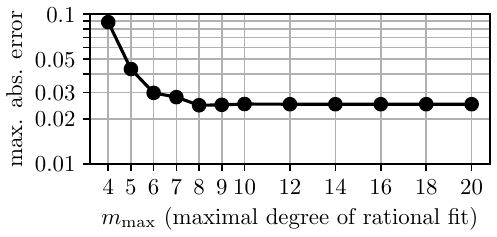}
    \label{fig:airy_err_aaa}
\end{subfigure}
\begin{subfigure}[t]{\linewidth}
    \caption{}
    \includegraphics[width=\linewidth]{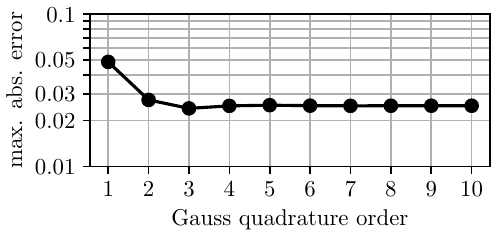}
    \label{fig:airy_err_quad}
\end{subfigure}
\caption{Maximal absolute error (across the spatial domain) between the MGO and exact solution for the Airy Problem when varying (a) the barycentric rational maximal interpolation degree and (b) the order of the Gauss-Freud quadrature. (a) used 10 quadrature points and (b) used $m_{\mathrm{max}}=20$.}
\label{fig:airy_err}
\end{figure}

\review{To quantify the error of the \ac{MGO} solution we show in FIG. \ref{fig:airy_err} the maximal absolute deviation between the MGO solution and exact Airy function for varying parameters. FIG. \ref{fig:airy_err_aaa} shows how the error depends on the maximal degree of the barycentric rational interpolation and \ref{fig:airy_err_quad} shows the dependence of the error on the number of quadrature points in the Gauss-Freud quadrature. In both figures the absolute error quickly converges to about 0.025 corresponding to a relative error of about 5\%. This order of magnitude error would only result in minor corrections to most subsequent calculations.} We note that the performance of this numerical implementation is comparable to the results from Ref. \cite{Donelly2021} where the exact analytical function $f_t(\epsilon)$ is known. Therefore, the error sources in our results are likely the same as for the analytic \ac{MGO} results previously shown in the literature. The deviations might be attributed to the limits of using Gauss-Freud quadrature near the caustic, the linear contour approximation or perhaps most significant the \ac{NIMT} approximation.

\review{
\begin{figure}[ht]
    \centering
    \includegraphics[width=\linewidth]{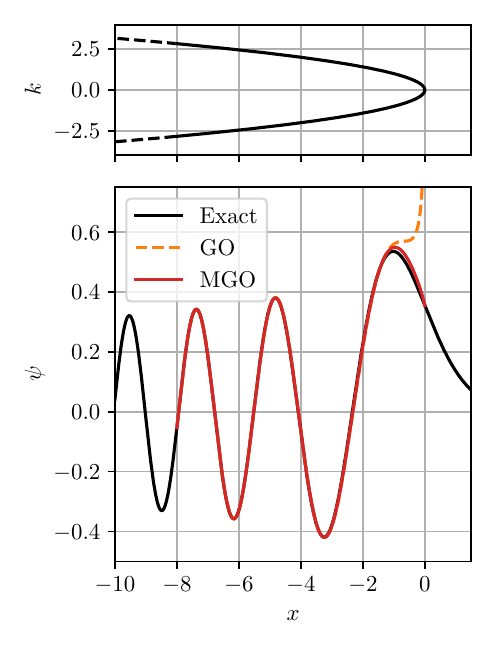}
    \caption{Solution to Airy's Equation, Eq. \eqref{eq:Ai_eq}. The top plot shows the ray phase space trajectory. The solid line is the ray trace used for calculating the final field. The dashed line indicates the extra tracing carried out to get sufficient data for calculating the field at the boundaries of the trace domain. In the bottom plot we see the resulting wave field. We have shown both the exact wave function, $\psi(x) = \Ai(x)$, the \ac{GO} approximation, and the solution from applying the \ac{MGO} algorithm.}
    \label{fig:airy}
\end{figure}
}

\subsection{Weber's Equation \label{sec:weber_results}}

\begin{figure}
    \centering
    \includegraphics[width=\linewidth]{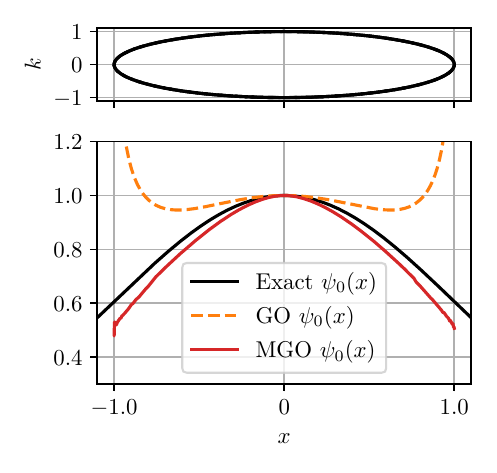}
    \caption{Numerical Solution to Weber's Equation, \ref{eq:We_eq}, for the ground state ($\nu=0$). On the top plot we see the ray phase space trajectory. The ray path is seen to be periodic in phase space, and the ray tracing was automatically stopped after 1 cycle. However the ray was traced a bit more than 1 cycle to have sufficient data on the boundaries. In the bottom, we see the reconstructed wave field in position space within the eikonal approximation of \ac{GO}. For comparison, we have also included the exact wave function solution given by \cref{eq:Webers_solution}.}
    \label{fig:weber_nu0}
\end{figure}

\begin{figure}
    \centering
    \includegraphics[width=\linewidth]{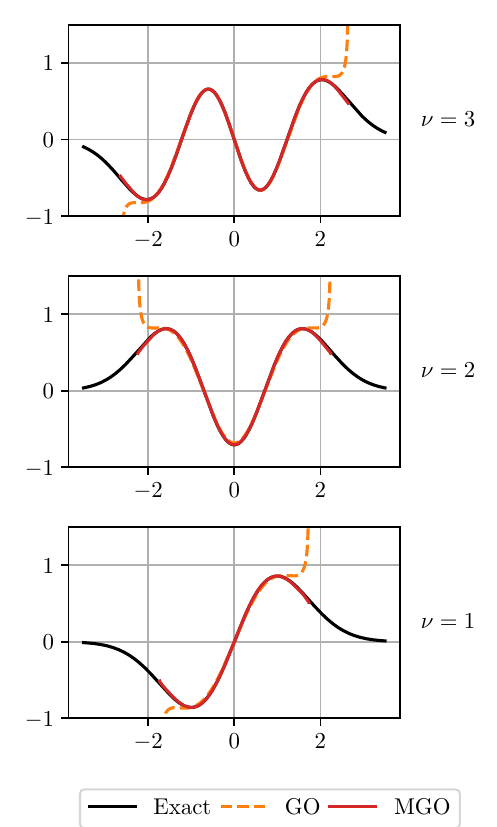}
    \caption{Numerical Solutions to Weber's Equation, Eq. \eqref{eq:Webers_solution}, for first three excited states $\nu=1, 2, 3$. See caption to FIG. \ref{fig:weber_nu0}.}
    \label{fig:weber_excited_modes}
\end{figure}

\begin{figure*}
    \centering
    \includegraphics[width=\textwidth]{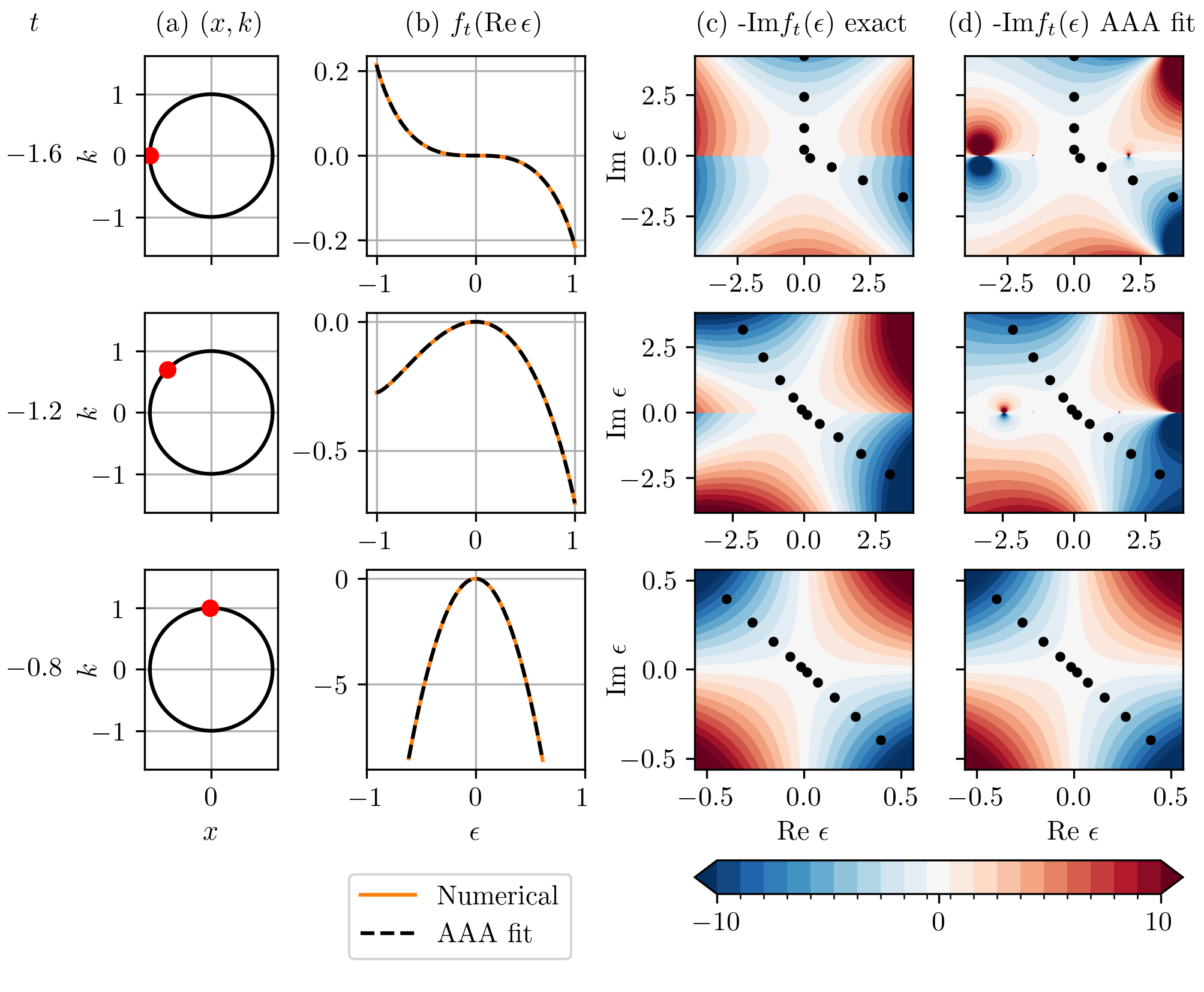}
    \caption{Inspection of barycentric rational interpolation needed for the analytic continuation in order to evaluate the steepest descent integral as part of the inverse metaplectic transform step for the Weber equation example. (a) Ray manifold with red dot indicating phase space location at time $t$, (b) Metaplectically transformed phase factor, $f_t$, as functions of $\epsilon = X_t - X_t(t) \in \RR$, (c) exact "$-$Im" part of the analytically continued phase factor from Eq. \eqref{eq:Webers_ft} evaluated on $\epsilon \in \CC$, (d) barycentric rational interpolation of $f_t$ evaluated on $\epsilon \in \CC$. In (b) we have included the numerical result and the barycentric rational interpolation. For (c) and (d) we have also shown the Gauss quadrature node loci ($\epsilon_j = \lambda l_j e^{\sg_\pm}$) along the steepest descent contour used to evaluate the integral of the inverse metaplectic transform.}
    \label{fig:weber_steps_2}
\end{figure*}

\noindent As our next example we consider Weber's equation:
\begin{gather}
\label{eq:We_eq}
    (2E_\nu + \p_x^2-x^2)\psi(x)=0,
\end{gather}
where $E_\nu=\nu+1/2$ and $\nu\in\NN_0$. Weber's equation describes the quantum harmonic oscillator, with $E_\nu$ being the energy associated with the mode number $\nu$. In a plasma physics context, this could be an O-mode inside a non-monotonic density profile, where it meets a cutoff on either side of the peak. These cutoffs are caustics and are found at $x=\pm\sqrt{2E_\nu}$. After a Wigner-Weyl transform, the dispersion symbol is found to be (Ref. \cite{Lopez2022}):
\begin{gather}
    \Dc(x,k) = (2E_\nu-k^2-x^2) = 0.
\end{gather}
An exact solution can be expressed in terms of the Hermite polynomials, $H_\nu(x)$, as
\begin{gather}
    \psi_\nu(x)=\dfrac{\pi^{-1/4}}{\sqrt{2^\nu\nu!}}H_\nu(x)e^{-x^2/2},\label{eq:Webers_solution}
\end{gather}
Again, we also cite the analytical results from Ref. \cite{Lopez2020,Lopez2022} of the function for the contour integral which we later compare to in FIG. \ref{fig:weber_steps_2}:
\begin{equation}
\label{eq:Webers_ft}
\begin{gathered}
    f_t(\epsilon)=\dfrac{\epsilon}{2}\sqrt{2E_\nu-\epsilon^2}+E_\nu\sin^{-1}\left( \dfrac{\epsilon}{\sqrt{2E_\nu}}\right) \\
    -\; \dfrac{\tan(2t)}{2}\epsilon^2-\sqrt{2E_\nu}\epsilon.
\end{gathered}
\end{equation}
For the numerical solutions, we follow a very similar procedure to Airy's equation. This time, we initiate the ray tracing at $x=x_0=-R_{\nu}$, since $R_\nu = \sqrt{2 E_{\nu}}$ is the radius of the oscillation. The initial condition is chosen to match the exact solution at $x=x_0$. We use a "ghost margin" of 16 \% such that we trace 16 \% of the number of time points at each side of the ray trace for ghost points. We show the result of the fundamental mode in FIG. \ref{fig:weber_nu0}. Again, the \ac{GO} solution diverges at the caustics but here it only agrees well with the exact solution close to $x=0$, in the middle between the two caustics. The \ac{MGO} solution generally agrees much better with the exact solution but is visibly less accurate than for the Airy equation. Still, the solution stays within $\sim 10\%$ of the exact solution at all points. For higher mode numbers shown in FIG. \ref{fig:weber_excited_modes}, there is a better agreement between the \ac{MGO} solution and the exact solution. In phase space, the modes form closed circles of radius $\sqrt{2\nu+1}$. The higher modes therefore have a smaller curvature, meaning that the frame is rotated slower with respect to $t$. When we inspect the barycentric rational AAA interpolations and the reconstructed $-\mathrm{Im}f_t(\epsilon)$ in FIG. \ref{fig:weber_steps_2}, we see that although the AAA fit is excellent inside the $\epsilon$ domain, where there is data, the contour integral uses quadrature points which extend further away from the origin than the data points. Outside the domain, we see the occurrence of what appears to be Froissart Doublets, which can be thought of as spurious pole-zero pairs very close together such that they nearly cancel, Ref. \cite{Nakatsukasa2018}. In both the purely numerical case as well as the case where $f_t(\epsilon)$ is known, the straight line contour approximation and the algorithm that determines the directions appear to mostly work well. However, some of the points furthest from the origin seem to deviate slightly from the true steepest descent contour. Still, the main contribution to the Gauss-Freud quadrature comes from the points closest to the origin where the function is reconstructed very well, and for this reason, the final result comes close to the exact solution. We note that analytical results using the \ac{MGO} method in Ref. \cite{Lopez2020} perform comparably well.

\subsection{X-B mode coupling in ASDEX Upgrade \label{sec:XB_results}}
\noindent \review{In the final example, we apply the code to the problem of X-B coupling where X-mode couples to an \ac{EBW} at the upper hybrid layer in a hot magnetized plasma. In this phenomenon a forward propagating electromagnetic wave is coupled to a backward propagating electrostatic wave. X-B mode coupling occurs when generating \acp{EBW} for heating and current drive (Ref. \cite{Laqua1997, Guo2022}) and for wave trapping related to low threshold two-plasmon decay instabilities (Ref. \cite{Clod2024}). The nomenclature of what is meant by X-B and the different waves vary across the literature. We refer to the forward propagating wave as X-mode, the backward propagating wave as the \ac{EBW} and the turning point as the upper hybrid layer. Unlike the cold O- and X-modes, the X-mode and the \ac{EBW} are described by the same dispersion symbol and are simply two different parts of the same ray trajectory with different physical wave characteristics.} The turning point is a fold caustic and the phase space trajectory around this point is similar to that of the cutoff in Airy's equation except that the caustic occurs at non-zero $k$ and the in- and outgoing branches describe different types of waves. This problem has not previously been treated analytically with \ac{MGO} and is unlikely to ever be as the kinetic dispersion relations for magnetized plasmas are rather complicated. In place of a comparison with analytical theory, we will make use of 1D fully kinetic \ac{PIC} simulations of X-B coupling at the upper hybrid layer. \review{We take parameters inspired by the 
\ac{ASDEX} which is a medium sized tokamak equipped with several gyrotrons used for electron cyclotron resonance heating and current drive as well as for collective Thomson scattering diagnostics, Refs. \cite{Kudlacek2024, Hansen2019b}. For these parameters the upper hybrid layer of the gyrotrons is found between the fundamental and second harmonic electron cyclotron frequency.}

\begin{figure}
    \centering
    \includegraphics[width=\linewidth]{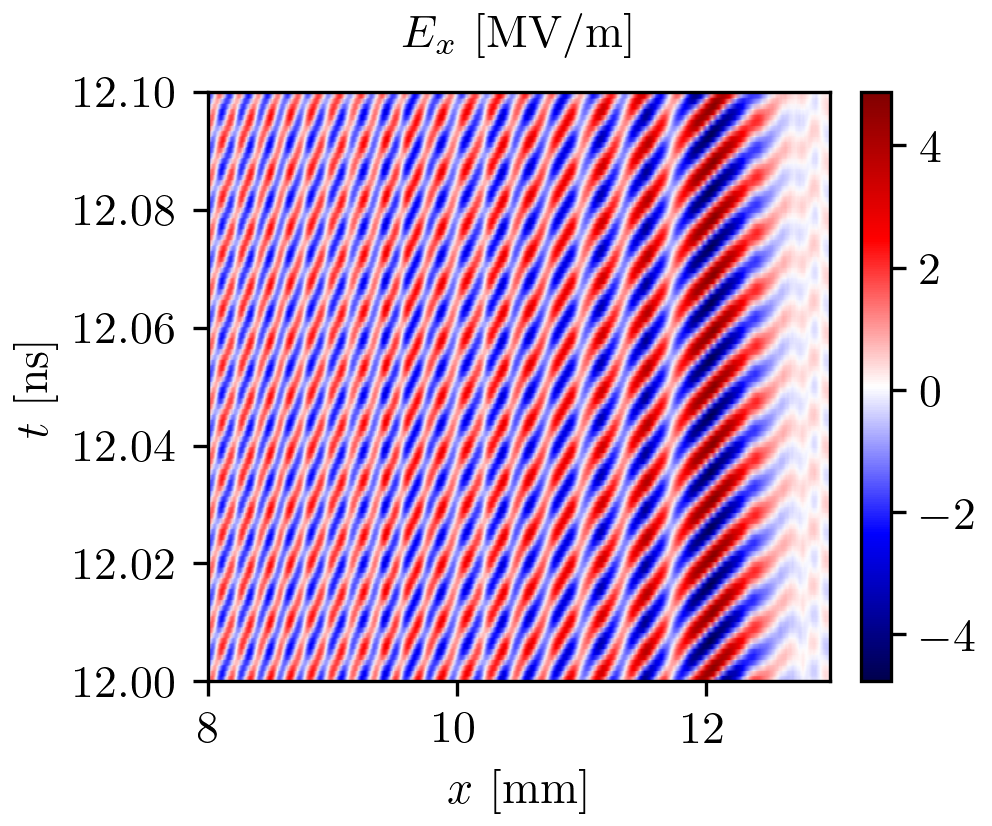}
    \caption{Longitudinal electric field, $E_x$, in 1D PIC simulations of X-B mode coupling in \ac{ASDEX}. An X-mode wave is excited at $x=$ 0 mm and propagates to the upper hybrid layer near $x=$ 12 mm where it couples to a backward propagating electron Bernstein wave. At the shown time, the returning electron Bernstein wave has made it back to the left boundary of the figure and an interference pattern can be seen.}
    \label{fig:EPOCHEx}
\end{figure}

For the \ac{PIC} simulations, we use the \ac{PIC} code EPOCH, Ref. \cite{Arber2015}. It is a low power simulation that has previously been used in a study of parametric decay in \ac{ASDEX}, Ref. \cite{Senstius2020b}, with 1 spatial and 3 velocity dimensions. It is a deuterium plasma with a linear density profile of $n_{\mathrm{e}}(x)=n_{\mathrm{i}}(x)=5.4\times10^{19}\,\mathrm{m}^{-3} - x \, 2.0\times 10^{21}\,\mathrm{mm}^{-4}$, where subscript e and i refer to electrons and deuterons, with a uniform temperature profile of $T_{\mathrm{e}}=T_{\mathrm{i}}=300$ eV and a magnetic field of $B=3.35$ T, pointing perpendicular to the $x$-direction. The domain is 0 mm $\leq x\leq $ 13.5 mm with 1666 grid points and $6\times10^4$ macroparticles per grid point, corresponding approximately to 64 gridpoints per wavelength at the upper hybrid layer. The particle boundary conditions replace lost particles at the boundaries with new thermally distributed ones. The field boundary conditions are open with a $\omega/(2\pi)=105$ GHz X-mode polarized wave of intensity $I=7\times10^7\;\mathrm{W/m^2}$ at the $x = 0$ boundary. With these parameters, the upper hybrid layer is found in the vicinity of $x=$ 12 mm and an excerpt of the longitudinal electric field can be seen in FIG. \ref{fig:EPOCHEx}. An interference pattern in space can be seen in the figure and the field is seen to peak at a finite value close to $x=$ 12 mm. The field appears to have a simple harmonic time dependence on the shown time scale.

To model this with \ac{GO} and \ac{MGO}, we use the following dispersion relation (Refs. \cite{Senstius2024,Crawford1965,Hansen2019})
\begin{gather}
\label{eq:XB_disp}
    \mathcal{D}(x,k,\omega)=K_1(x,k,\omega)k^2-\left(\dfrac{\omega}{c}\right)^2\left(S^2-D^2\right)\\
K_1(x,k,\omega)
= 1 + 
\dfrac{\omega_{\mathrm{pe}}^2}{\omega_{\mathrm{ce}}^2}
\exp(-\lambda)
\notag\\
\times\int_0^\pi \review{\dd{\psi}}\dfrac{\sin\left(\psi\left(\dfrac{\omega}{\omega_{\mathrm{ce}}}\right)\right)\sin(\psi)\exp(-\lambda\cos(\psi))}{\sin\left(\pi\left(\dfrac{\omega}{\omega_{\mathrm{ce}}}\right)\right)},
\end{gather}
where $S=1-\omega_{\mathrm{pe}}^2/(\omega^2-\omega_{\mathrm{ce}}^2)$ and $D=\omega_{\mathrm{ce}}\omega_{\mathrm{pe}}^2/(\omega(\omega^2-\omega_{\mathrm{ce}}^2))$ are the Stix sum and difference parameters, $\lambda=k^2v_{\mathrm{Te}}^2/(2\omega_{\mathrm{ce}}^2)$ is a normalized squared wavenumber, $\omega_{\mathrm{ce}}=eB/m_{\mathrm{e}}$ and $\omega_{\mathrm{pe}}=\sqrt{e^2n_e/(\epsilon_0m_{\mathrm{e}})}$ are the electron cyclotron and plasma frequencies, and $v_{\mathrm{Te}}=\sqrt{2T_{\mathrm{e}}/m_{\mathrm{e}}}$ is the electron thermal speed. Note that we are neglecting contributions from the ions because we are considering a wave in the electron frequency range and the large mass ratio renders the ion contributions insignificant. As has been reported earlier in the literature, the theoretical dispersion relation does not perfectly match what is observed in EPOCH \cite{Senstius2021}. The reasons could include numerical dispersion and that non-local and nonlinear effects are neglected in the linear dispersion relation but not in EPOCH.

\begin{figure}
    \centering
    \includegraphics[width=\linewidth]{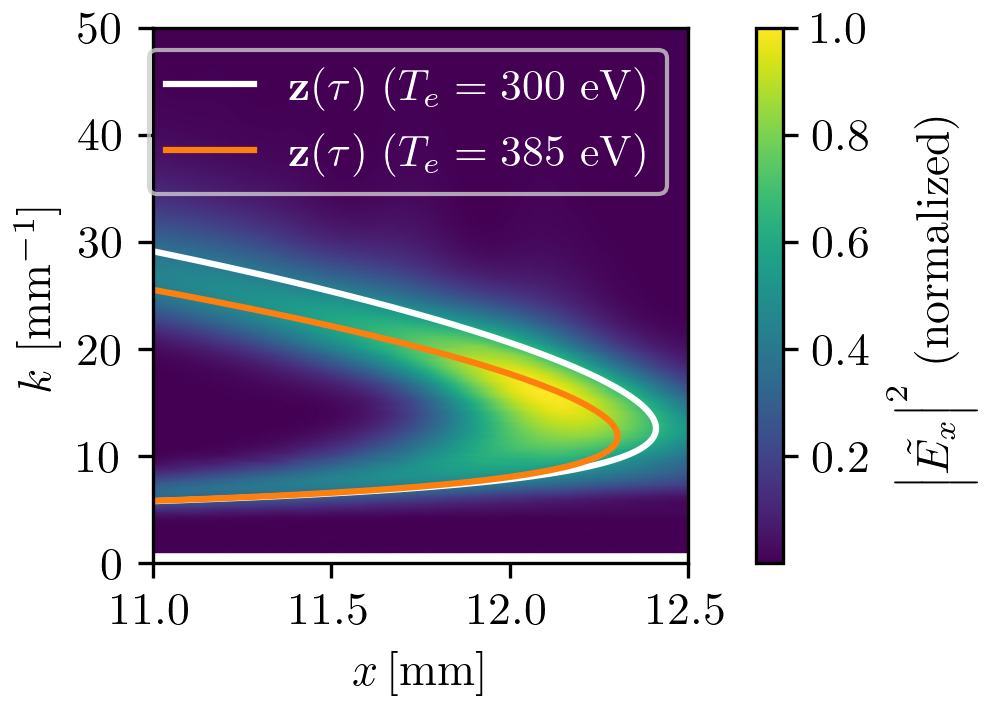}
    \caption{Ray phase space trajectory for X-B mode coupling plotted on top of normalized spectral density from PIC simulation. The colored spectral density plot shows the normalized norm square of the continuous wavelet transform of $E_x$ from a PIC simulation (see FIG. \ref{fig:EPOCHEx}) at a specific time $(t = 8 \, \mathrm{ns})$ using a complex Morlet wavelet. Meanwhile the white curve shows the ray trace obtained from the dispersion relation in Eq. \eqref{eq:XB_disp} with an electron temperature of $T_e = 300 \, \mathrm{eV}$. The \review{orange} curve shows the trace if instead $T_e = 385 \, \mathrm{eV}$.}
    \label{fig:XB_trace}
\end{figure}

To get a better matching dispersion curve, we multiply the electron temperature by a factor of 1.28 in our \ac{GO} and \ac{MGO} calculations. This temperature factor was found by varying the temperature such that the absolute difference between the \ac{GO} and \ac{PIC} fields away from the caustic was minimized. \review{The different ray traces corresponding to an electron temperature of $300 \, \mathrm{eV}$ and $1.28 \times 300 \approx 385 \, \mathrm{eV}$ are shown on \cref{fig:XB_trace} on top of the spectral density from the \ac{PIC} simulation. Clearly, neither of the ray traces match the numerical simulation exactly, but the curve with higher electron temperature goes through the high intensity region in phase space and should therefore match better with the \ac{PIC} simulation.}

Similar to Airy's equation, we trace a ray starting as an X-mode at $x=0$ and end the tracing when the returning \ac{EBW} reaches $x=0$. We then reconstruct the \ac{MGO} wave field and define the phase and amplitude such that the absolute difference between the \ac{MGO}/\ac{GO} and \ac{PIC} fields is minimized across the domain $x \in [11 \, \mathrm{mm}, 11.5 \, \mathrm{mm}]$. As an alternative, one could have matched the final \review{\ac{MGO}/\ac{GO}} fields to the PIC simulation at a specific $x$-location similar to the procedure of the Airy and Weber examples. However this alternative approach would be very sensitive to the specific $x$-location chosen since, as we shall see, we have a rapidly oscillating electric field and there is not exact phase match between the PIC field and the \ac{MGO}/\ac{GO} fields.

\begin{figure}
    \centering
    \includegraphics[width=\linewidth]{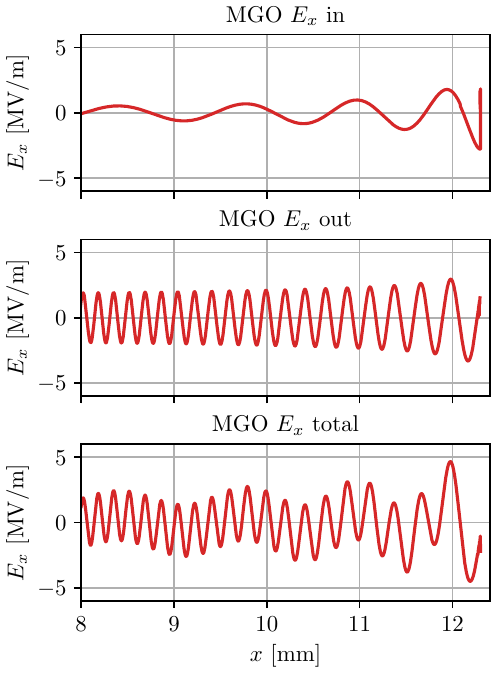}
    \caption{Plots of the ingoing (top plot), outgoing (center plot) and combined (bottom plot) field obtained from \ac{MGO} for the case of X-B mode coupling with $T_e = 385 \, \mathrm{eV}$.}
    \label{fig:XB_in_and_out}
\end{figure}

On FIG. \ref{fig:XB_in_and_out} we show the resulting incoming and outgoing and combined wave fields according to the \ac{MGO} implementation.
Crucially, we note that the \ac{MGO} solution is finite everywhere and looks smooth except right before the turning point in the right side, where the curve becomes a little noisy. We attribute this noise to the interpolation based on the \ac{AAA} algorithm in combination with the straight contour approximation which leads the contours close to spurious poles of the \ac{AAA} extrapolation.

\begin{figure}
    \centering
    \includegraphics[width=\linewidth]{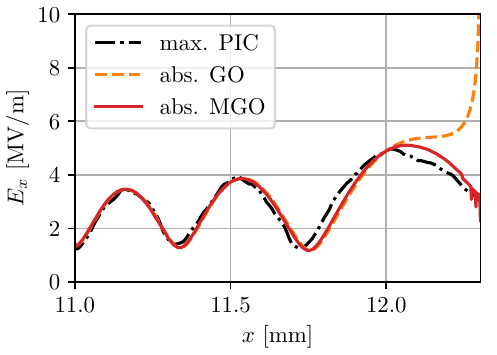}
    \caption{A comparison of the envelope of the longitudinal electric wave field for X-B mode coupling. The absolute value of the \ac{GO}/\ac{MGO} fields (with $T_e = 385 \, \mathrm{eV}$) are compared to the maximum of the \ac{PIC} result (the \ac{PIC} field is maximized over a time window of about 5 ns).
    The amplitude of the \ac{GO} and \ac{MGO} solutions were set by minimizing the absolute difference between the absolute MGO field and the maximum \ac{PIC} field over the interval $x \in \qty[11 \, \mathrm{mm}, 11.5 \, \mathrm{mm}]$.}
    \label{fig:XBFinalResult}
\end{figure}

Finally, we compare the amplitude of the complex envelope function of both the \ac{GO} and \ac{MGO} solutions with that of the PIC simulations. Since the fields are real-valued in the PIC simulations we determine the envelope as the maximum in the time interval $6.5 \, \mathrm{ns} < t <12 \, \mathrm{ns}$ at each grid point inside the domain. No clear transients occur near the upper hybrid layer in this period. The resulting comparison is shown in FIG. \ref{fig:XBFinalResult}. All solutions agree well on shape and magnitude with some minor deviations, except for the \ac{GO} solution which diverges at the upper hybrid layer. It is worth noting that the PIC simulations model many more effects than the simple linear dispersion relation is capturing. As mentioned earlier, non-local and nonlinear effects are neglected even though the upper hybrid layer is often associated with nonlinear effects due, in part, to the caustic. Furthermore, although both the X-mode and electron Bernstein wave become approximately electrostatic near the upper hybrid layer, the X-B mode coupling is not truly a scalar wave problem. Still, our numerical implementation of \ac{MGO} is capable of capturing the features of the PIC wave field, provided that the dispersion manifold can be determined sufficiently well.

\section{Conclusion and discussion}
\noindent We have presented the first unsupervised numerical implementation of \ac{MGO}, relying only on a 1D discrete phase space trajectory produced by a ray tracer. The code needs no additional information to the ray tracer and reconstructs the wave field with good agreement, and importantly without singularities, for the bench marking examples of Airy's and Weber's equation. The numerical solutions display minor deviations near the caustics with a relative error within 10 \% in the worst case example in Weber's equation. This discrepancy may be attributed to the numerical integration method used to evaluate the inverse metaplectic transform or to an approximation of the \ac{NIMT}. The contour integrals are performed using Gauss-Freud quadrature and a barycentric rational function fit, which reconstructs the target functions with great precision and does a decent job extrapolating to the complex domain. A problem in the current implementation is that the barycentric rational interpolation suffers from Froissart doublets. An approach to remove these doublets is proposed in Ref. \cite{Nakatsukasa2018}, but not yet implemented in the numerical implementation presented here. Though there is room for optimization, our numerical implementation reconstructs the wave field in seconds to a few minutes on a modern laptop (MacBook Pro 2019) depending on the resolution needed for the problem (about 30 seconds for Airy's equation with 700 ray tracing points and 1.5 minutes for the X-B example with 2500 ray tracing points). The numerical implementation is fully automated \review{and the implementation can in principle be extended to more spatial dimensions in future work. However, it remains an unresolved problem to generalize the steepest-descent methods used in 1D in this work to multidimensional integrals.} Importantly, the \ac{MGO} algorithm is agnostic to the type of linear wave equation, though the current formulation of \ac{MGO} is reserved to time-invariant scalar waves. 

When applied to the problem of X-B mode coupling in a magnetized fusion plasma, the numerical implementation of \ac{MGO} is capable of achieving good agreement with \ac{PIC} simulations, with similar or better performance than \ac{GO} and crucially without the divergent behavior at the caustic. This opens up for applications in reduced modeling of phenomena which depend nonlinearly on the electric field intensity such as three-wave interactions and stochastic heating.

\section{Acknowledgments}
We thank Dr. Nicolas Lopez for fruitful discussions. This work has been carried out within the framework of the EUROfusion Consortium, funded by the European Union via the Euratom Research and Training Programme (Grant Agreement No 101052200 — EUROfusion). Views and opinions expressed are however those of the author(s) only and do not necessarily reflect those of the European Union or the European Commission. Neither the European Union nor the European Commission can be held responsible for them. The work presented here is supported by the Carlsberg Foundation, grant CF23-0181, and the Novo Nordisk Foundation, NNF22OC0076017.

\section{References}
\bibliography{references.bib}

\begin{thebibliography}{48}%
\makeatletter
\providecommand \@ifxundefined [1]{%
 \@ifx{#1\undefined}
}%
\providecommand \@ifnum [1]{%
 \ifnum #1\expandafter \@firstoftwo
 \else \expandafter \@secondoftwo
 \fi
}%
\providecommand \@ifx [1]{%
 \ifx #1\expandafter \@firstoftwo
 \else \expandafter \@secondoftwo
 \fi
}%
\providecommand \natexlab [1]{#1}%
\providecommand \enquote  [1]{``#1''}%
\providecommand \bibnamefont  [1]{#1}%
\providecommand \bibfnamefont [1]{#1}%
\providecommand \citenamefont [1]{#1}%
\providecommand \href@noop [0]{\@secondoftwo}%
\providecommand \href [0]{\begingroup \@sanitize@url \@href}%
\providecommand \@href[1]{\@@startlink{#1}\@@href}%
\providecommand \@@href[1]{\endgroup#1\@@endlink}%
\providecommand \@sanitize@url [0]{\catcode `\\12\catcode `\$12\catcode `\&12\catcode `\#12\catcode `\^12\catcode `\_12\catcode `\%12\relax}%
\providecommand \@@startlink[1]{}%
\providecommand \@@endlink[0]{}%
\providecommand \url  [0]{\begingroup\@sanitize@url \@url }%
\providecommand \@url [1]{\endgroup\@href {#1}{\urlprefix }}%
\providecommand \urlprefix  [0]{URL }%
\providecommand \Eprint [0]{\href }%
\providecommand \doibase [0]{http://dx.doi.org/}%
\providecommand \selectlanguage [0]{\@gobble}%
\providecommand \bibinfo  [0]{\@secondoftwo}%
\providecommand \bibfield  [0]{\@secondoftwo}%
\providecommand \translation [1]{[#1]}%
\providecommand \BibitemOpen [0]{}%
\providecommand \bibitemStop [0]{}%
\providecommand \bibitemNoStop [0]{.\EOS\space}%
\providecommand \EOS [0]{\spacefactor3000\relax}%
\providecommand \BibitemShut  [1]{\csname bibitem#1\endcsname}%
\let\auto@bib@innerbib\@empty
\bibitem [{\citenamefont {H\o{}jlund~Marholt}\ \emph {et~al.}(2024)\citenamefont {H\o{}jlund~Marholt}, \citenamefont {Senstius},\ and\ \citenamefont {Nielsen}}]{Marholt2024}%
  \BibitemOpen
  \bibfield  {author} {\bibinfo {author} {\bibfnamefont {R.}~\bibnamefont {H\o{}jlund~Marholt}}, \bibinfo {author} {\bibfnamefont {M.~G.}\ \bibnamefont {Senstius}}, \ and\ \bibinfo {author} {\bibfnamefont {S.~K.}\ \bibnamefont {Nielsen}},\ }\href {\doibase 10.1103/PhysRevE.110.025208} {\bibfield  {journal} {\bibinfo  {journal} {Phys. Rev. E}\ }\textbf {\bibinfo {volume} {110}},\ \bibinfo {pages} {025208} (\bibinfo {year} {2024})}\BibitemShut {NoStop}%
\bibitem [{\citenamefont {Tracy}\ \emph {et~al.}(2012)\citenamefont {Tracy}, \citenamefont {Brizard}, \citenamefont {Richardson},\ and\ \citenamefont {Kaufman}}]{Tracy2012}%
  \BibitemOpen
  \bibfield  {author} {\bibinfo {author} {\bibfnamefont {E.~R.}\ \bibnamefont {Tracy}}, \bibinfo {author} {\bibfnamefont {A.~J.}\ \bibnamefont {Brizard}}, \bibinfo {author} {\bibfnamefont {A.~S.}\ \bibnamefont {Richardson}}, \ and\ \bibinfo {author} {\bibfnamefont {A.~N.}\ \bibnamefont {Kaufman}},\ }\href {\doibase 10.1017/CBO9780511667565} {\emph {\bibinfo {title} {Ray tracing and beyond: Phase space methods in Plasma wave theory}}}\ (\bibinfo  {publisher} {Cambridge University Press},\ \bibinfo {year} {2012})\BibitemShut {NoStop}%
\bibitem [{\citenamefont {Batchleor}\ \emph {et~al.}(1980)\citenamefont {Batchleor}, \citenamefont {Goldfinger},\ and\ \citenamefont {Weitzner}}]{Batchleor1980}%
  \BibitemOpen
  \bibfield  {author} {\bibinfo {author} {\bibfnamefont {D.~B.}\ \bibnamefont {Batchleor}}, \bibinfo {author} {\bibfnamefont {R.~C.}\ \bibnamefont {Goldfinger}}, \ and\ \bibinfo {author} {\bibfnamefont {H.}~\bibnamefont {Weitzner}},\ }\href {\doibase 10.1109/TPS.1980.4317276} {\bibfield  {journal} {\bibinfo  {journal} {IEEE Transactions on Plasma Science}\ }\textbf {\bibinfo {volume} {8}},\ \bibinfo {pages} {78} (\bibinfo {year} {1980})}\BibitemShut {NoStop}%
\bibitem [{\citenamefont {Donné}(1995)}]{Donne1995}%
  \BibitemOpen
  \bibfield  {author} {\bibinfo {author} {\bibfnamefont {A.~J.~H.}\ \bibnamefont {Donné}},\ }\href {\doibase 10.1063/1.1145516} {\bibfield  {journal} {\bibinfo  {journal} {Review of Scientific Instruments}\ }\textbf {\bibinfo {volume} {6}},\ \bibinfo {pages} {3407–} (\bibinfo {year} {1995})}\BibitemShut {NoStop}%
\bibitem [{\citenamefont {Nielsen}\ \emph {et~al.}(2008)\citenamefont {Nielsen}, \citenamefont {Bindslev}, \citenamefont {Porte}, \citenamefont {Hoekzema}, \citenamefont {Korsholm}, \citenamefont {Leipold}, \citenamefont {Meo}, \citenamefont {Michelsen}, \citenamefont {Michelsen}, \citenamefont {Oosterbeek}, \citenamefont {Tsakadze}, \citenamefont {Van~Wassenhove}, \citenamefont {Westerhof},\ and\ \citenamefont {Woskov}}]{Nielsen2008}%
  \BibitemOpen
  \bibfield  {author} {\bibinfo {author} {\bibfnamefont {S.~K.}\ \bibnamefont {Nielsen}}, \bibinfo {author} {\bibfnamefont {H.}~\bibnamefont {Bindslev}}, \bibinfo {author} {\bibfnamefont {L.}~\bibnamefont {Porte}}, \bibinfo {author} {\bibfnamefont {J.~A.}\ \bibnamefont {Hoekzema}}, \bibinfo {author} {\bibfnamefont {S.~B.}\ \bibnamefont {Korsholm}}, \bibinfo {author} {\bibfnamefont {F.}~\bibnamefont {Leipold}}, \bibinfo {author} {\bibfnamefont {F.}~\bibnamefont {Meo}}, \bibinfo {author} {\bibfnamefont {P.~K.}\ \bibnamefont {Michelsen}}, \bibinfo {author} {\bibfnamefont {S.}~\bibnamefont {Michelsen}}, \bibinfo {author} {\bibfnamefont {J.~W.}\ \bibnamefont {Oosterbeek}}, \bibinfo {author} {\bibfnamefont {E.~L.}\ \bibnamefont {Tsakadze}}, \bibinfo {author} {\bibfnamefont {G.}~\bibnamefont {Van~Wassenhove}}, \bibinfo {author} {\bibfnamefont {E.}~\bibnamefont {Westerhof}}, \ and\ \bibinfo {author} {\bibfnamefont {P.}~\bibnamefont {Woskov}},\ }\href {\doibase 10.1103/PhysRevE.77.016407} {\bibfield  {journal} {\bibinfo  {journal} {Phys. Rev. E}\ }\textbf {\bibinfo {volume} {77}},\ \bibinfo {pages} {016407} (\bibinfo {year} {2008})}\BibitemShut {NoStop}%
\bibitem [{\citenamefont {Prater}\ \emph {et~al.}(2008)\citenamefont {Prater}, \citenamefont {Farina}, \citenamefont {Gribov}, \citenamefont {Harvey}, \citenamefont {Ram}, \citenamefont {Lin-Liu}, \citenamefont {Poli}, \citenamefont {Smirnov}, \citenamefont {Volpe}, \citenamefont {Westerhof}, \citenamefont {Zvonkov},\ and\ \citenamefont {the Group}}]{Prater2008}%
  \BibitemOpen
  \bibfield  {author} {\bibinfo {author} {\bibfnamefont {R.}~\bibnamefont {Prater}}, \bibinfo {author} {\bibfnamefont {D.}~\bibnamefont {Farina}}, \bibinfo {author} {\bibfnamefont {Y.}~\bibnamefont {Gribov}}, \bibinfo {author} {\bibfnamefont {R.}~\bibnamefont {Harvey}}, \bibinfo {author} {\bibfnamefont {A.}~\bibnamefont {Ram}}, \bibinfo {author} {\bibfnamefont {Y.-R.}\ \bibnamefont {Lin-Liu}}, \bibinfo {author} {\bibfnamefont {E.}~\bibnamefont {Poli}}, \bibinfo {author} {\bibfnamefont {A.}~\bibnamefont {Smirnov}}, \bibinfo {author} {\bibfnamefont {F.}~\bibnamefont {Volpe}}, \bibinfo {author} {\bibfnamefont {E.}~\bibnamefont {Westerhof}}, \bibinfo {author} {\bibfnamefont {A.}~\bibnamefont {Zvonkov}}, \ and\ \bibinfo {author} {\bibfnamefont {I.~S. S. O.~T.}\ \bibnamefont {the Group}},\ }\href {\doibase 10.1088/0029-5515/48/3/035006} {\bibfield  {journal} {\bibinfo  {journal} {Nuclear Fusion}\ }\textbf {\bibinfo {volume} {48}},\ \bibinfo {pages} {035006} (\bibinfo {year} {2008})}\BibitemShut {NoStop}%
\bibitem [{\citenamefont {Nagasaki}\ \emph {et~al.}(2005)\citenamefont {Nagasaki}, \citenamefont {Isayama}, \citenamefont {Hayashi}, \citenamefont {Ozeki}, \citenamefont {Takechi}, \citenamefont {Oyama}, \citenamefont {Ide}, \citenamefont {Yamamoto},\ and\ \citenamefont {the JT-60~Team}}]{Nagasaki2005}%
  \BibitemOpen
  \bibfield  {author} {\bibinfo {author} {\bibfnamefont {K.}~\bibnamefont {Nagasaki}}, \bibinfo {author} {\bibfnamefont {A.}~\bibnamefont {Isayama}}, \bibinfo {author} {\bibfnamefont {N.}~\bibnamefont {Hayashi}}, \bibinfo {author} {\bibfnamefont {T.}~\bibnamefont {Ozeki}}, \bibinfo {author} {\bibfnamefont {M.}~\bibnamefont {Takechi}}, \bibinfo {author} {\bibfnamefont {N.}~\bibnamefont {Oyama}}, \bibinfo {author} {\bibfnamefont {S.}~\bibnamefont {Ide}}, \bibinfo {author} {\bibfnamefont {S.}~\bibnamefont {Yamamoto}}, \ and\ \bibinfo {author} {\bibnamefont {the JT-60~Team}},\ }\href {\doibase 10.1088/0029-5515/45/12/016} {\bibfield  {journal} {\bibinfo  {journal} {Nuclear Fusion}\ }\textbf {\bibinfo {volume} {45}},\ \bibinfo {pages} {1608} (\bibinfo {year} {2005})}\BibitemShut {NoStop}%
\bibitem [{\citenamefont {Preinhaelter}\ and\ \citenamefont {Kopecký}(1973)}]{Preinhaelter1973}%
  \BibitemOpen
  \bibfield  {author} {\bibinfo {author} {\bibfnamefont {J.}~\bibnamefont {Preinhaelter}}\ and\ \bibinfo {author} {\bibfnamefont {V.}~\bibnamefont {Kopecký}},\ }\href {\doibase 10.1017/S0022377800007649} {\bibfield  {journal} {\bibinfo  {journal} {Journal of plasma physics}\ }\textbf {\bibinfo {volume} {10}},\ \bibinfo {pages} {1} (\bibinfo {year} {1973})}\BibitemShut {NoStop}%
\bibitem [{\citenamefont {Ram}\ and\ \citenamefont {Schultz}(2000)}]{Ram2000}%
  \BibitemOpen
  \bibfield  {author} {\bibinfo {author} {\bibfnamefont {A.~K.}\ \bibnamefont {Ram}}\ and\ \bibinfo {author} {\bibfnamefont {S.~D.}\ \bibnamefont {Schultz}},\ }\href {\doibase 10.1063/1.1289689} {\bibfield  {journal} {\bibinfo  {journal} {Physics of plasmas}\ }\textbf {\bibinfo {volume} {7}},\ \bibinfo {pages} {4084} (\bibinfo {year} {2000})}\BibitemShut {NoStop}%
\bibitem [{\citenamefont {White}\ and\ \citenamefont {Chen}(1974)}]{White1974}%
  \BibitemOpen
  \bibfield  {author} {\bibinfo {author} {\bibfnamefont {R.~B.}\ \bibnamefont {White}}\ and\ \bibinfo {author} {\bibfnamefont {F.~F.}\ \bibnamefont {Chen}},\ }\href {\doibase 10.1088/0032-1028/16/7/002} {\bibfield  {journal} {\bibinfo  {journal} {Plasma physics}\ }\textbf {\bibinfo {volume} {16}},\ \bibinfo {pages} {565} (\bibinfo {year} {1974})}\BibitemShut {NoStop}%
\bibitem [{\citenamefont {Karney}\ and\ \citenamefont {Bers}(1977)}]{Karney1977}%
  \BibitemOpen
  \bibfield  {author} {\bibinfo {author} {\bibfnamefont {C.~F.~F.}\ \bibnamefont {Karney}}\ and\ \bibinfo {author} {\bibfnamefont {A.}~\bibnamefont {Bers}},\ }\href {\doibase 10.1103/PhysRevLett.39.550} {\bibfield  {journal} {\bibinfo  {journal} {Physical review letters}\ }\textbf {\bibinfo {volume} {39}},\ \bibinfo {pages} {550} (\bibinfo {year} {1977})}\BibitemShut {NoStop}%
\bibitem [{\citenamefont {Porkolab}\ and\ \citenamefont {Chang}(1978)}]{Porkolab1978}%
  \BibitemOpen
  \bibfield  {author} {\bibinfo {author} {\bibfnamefont {M.}~\bibnamefont {Porkolab}}\ and\ \bibinfo {author} {\bibfnamefont {R.~P.~H.}\ \bibnamefont {Chang}},\ }\href {\doibase 10.1103/RevModPhys.50.745} {\bibfield  {journal} {\bibinfo  {journal} {Reviews of modern physics}\ }\textbf {\bibinfo {volume} {50}},\ \bibinfo {pages} {745} (\bibinfo {year} {1978})}\BibitemShut {NoStop}%
\bibitem [{\citenamefont {Aleynikov}\ and\ \citenamefont {Marushchenko}(2019)}]{Aleynikov2019}%
  \BibitemOpen
  \bibfield  {author} {\bibinfo {author} {\bibfnamefont {P.}~\bibnamefont {Aleynikov}}\ and\ \bibinfo {author} {\bibfnamefont {N.~B.}\ \bibnamefont {Marushchenko}},\ }\href {\doibase 10.1016/j.cpc.2019.03.017} {\bibfield  {journal} {\bibinfo  {journal} {Computer Physics Communications}\ }\textbf {\bibinfo {volume} {241}},\ \bibinfo {pages} {40} (\bibinfo {year} {2019})}\BibitemShut {NoStop}%
\bibitem [{\citenamefont {Köhn-Seemann}\ \emph {et~al.}(2023)\citenamefont {Köhn-Seemann}, \citenamefont {Eliasson}, \citenamefont {Freethy}, \citenamefont {Holland},\ and\ \citenamefont {Vann}}]{Kohn-Seemann2023}%
  \BibitemOpen
  \bibfield  {author} {\bibinfo {author} {\bibfnamefont {A.}~\bibnamefont {Köhn-Seemann}}, \bibinfo {author} {\bibfnamefont {B.~E.}\ \bibnamefont {Eliasson}}, \bibinfo {author} {\bibfnamefont {S.~J.}\ \bibnamefont {Freethy}}, \bibinfo {author} {\bibfnamefont {L.~A.}\ \bibnamefont {Holland}}, \ and\ \bibinfo {author} {\bibfnamefont {R.~G.}\ \bibnamefont {Vann}},\ }\href {\doibase 10.1051/epjconf/202327701010} {\bibfield  {journal} {\bibinfo  {journal} {EPJ Web of Conferences}\ }\textbf {\bibinfo {volume} {277}},\ \bibinfo {pages} {01010} (\bibinfo {year} {2023})}\BibitemShut {NoStop}%
\bibitem [{\citenamefont {Lopez}\ and\ \citenamefont {Dodin}(2019)}]{Lopez2019}%
  \BibitemOpen
  \bibfield  {author} {\bibinfo {author} {\bibfnamefont {N.~A.}\ \bibnamefont {Lopez}}\ and\ \bibinfo {author} {\bibfnamefont {I.~Y.}\ \bibnamefont {Dodin}},\ }\href {\doibase 10.1364/josaa.36.001846} {\bibfield  {journal} {\bibinfo  {journal} {Journal of the Optical Society of America A}\ }\textbf {\bibinfo {volume} {36}} (\bibinfo {year} {2019}),\ 10.1364/josaa.36.001846}\BibitemShut {NoStop}%
\bibitem [{\citenamefont {Lopez}\ and\ \citenamefont {Dodin}(2020)}]{Lopez2020}%
  \BibitemOpen
  \bibfield  {author} {\bibinfo {author} {\bibfnamefont {N.~A.}\ \bibnamefont {Lopez}}\ and\ \bibinfo {author} {\bibfnamefont {I.~Y.}\ \bibnamefont {Dodin}},\ }\href {\doibase 10.1088/1367-2630/aba91a} {\bibfield  {journal} {\bibinfo  {journal} {New Journal of Physics}\ }\textbf {\bibinfo {volume} {22}},\ \bibinfo {pages} {083078} (\bibinfo {year} {2020})}\BibitemShut {NoStop}%
\bibitem [{\citenamefont {Lopez}\ and\ \citenamefont {Dodin}(2021)}]{Lopez2021}%
  \BibitemOpen
  \bibfield  {author} {\bibinfo {author} {\bibfnamefont {N.~A.}\ \bibnamefont {Lopez}}\ and\ \bibinfo {author} {\bibfnamefont {I.~Y.}\ \bibnamefont {Dodin}},\ }\href {\doibase 10.1364/JOSAA.417412} {\bibfield  {journal} {\bibinfo  {journal} {Journal of the Optical Society of America A}\ }\textbf {\bibinfo {volume} {38}} (\bibinfo {year} {2021}),\ 10.1364/JOSAA.417412}\BibitemShut {NoStop}%
\bibitem [{\citenamefont {Donnelly}\ \emph {et~al.}(2021)\citenamefont {Donnelly}, \citenamefont {Lopez},\ and\ \citenamefont {Dodin}}]{Donelly2021}%
  \BibitemOpen
  \bibfield  {author} {\bibinfo {author} {\bibfnamefont {S.~M.}\ \bibnamefont {Donnelly}}, \bibinfo {author} {\bibfnamefont {N.~A.}\ \bibnamefont {Lopez}}, \ and\ \bibinfo {author} {\bibfnamefont {I.~Y.}\ \bibnamefont {Dodin}},\ }\href {\doibase 10.1103/PhysRevE.104.025304} {\bibfield  {journal} {\bibinfo  {journal} {Physical Review E}\ }\textbf {\bibinfo {volume} {104}} (\bibinfo {year} {2021}),\ 10.1103/PhysRevE.104.025304}\BibitemShut {NoStop}%
\bibitem [{\citenamefont {{Nicolas Alexander Lopez}}(2022)}]{Lopez2022}%
  \BibitemOpen
  \bibfield  {author} {\bibinfo {author} {\bibnamefont {{Nicolas Alexander Lopez}}},\ }\emph {\bibinfo {title} {{Metaplectic geometrical optics}}},\ \href {\doibase 10.48550/arXiv.2210.03188} {Ph.D. thesis},\ \bibinfo  {school} {Princeton University} (\bibinfo {year} {2022})\BibitemShut {NoStop}%
\bibitem [{\citenamefont {Lopez}\ and\ \citenamefont {Dodin}(2022)}]{Lopez2022b}%
  \BibitemOpen
  \bibfield  {author} {\bibinfo {author} {\bibfnamefont {N.~A.}\ \bibnamefont {Lopez}}\ and\ \bibinfo {author} {\bibfnamefont {I.~Y.}\ \bibnamefont {Dodin}},\ }\href {\doibase 10.1063/5.0082241} {\bibfield  {journal} {\bibinfo  {journal} {Physics of Plasmas}\ }\textbf {\bibinfo {volume} {29}} (\bibinfo {year} {2022}),\ 10.1063/5.0082241}\BibitemShut {NoStop}%
\bibitem [{\citenamefont {Maslov}\ and\ \citenamefont {Fedoriuk}(1981)}]{Maslov1981}%
  \BibitemOpen
  \bibfield  {author} {\bibinfo {author} {\bibfnamefont {V.}~\bibnamefont {Maslov}}\ and\ \bibinfo {author} {\bibfnamefont {M.}~\bibnamefont {Fedoriuk}},\ }\href@noop {} {\emph {\bibinfo {title} {{Semi-Classical Approximation in Quantum : Mechanics.}}}},\ Mathematical Physics and Applied Mathematics ; 7.\ (\bibinfo  {publisher} {D. Reidel Publ. Co.},\ \bibinfo {year} {1981})\BibitemShut {NoStop}%
\bibitem [{\citenamefont {Keller}(1985)}]{Keller1985}%
  \BibitemOpen
  \bibfield  {author} {\bibinfo {author} {\bibfnamefont {J.~B.}\ \bibnamefont {Keller}},\ }\href {\doibase 10.1137/1027139} {\bibfield  {journal} {\bibinfo  {journal} {SIAM review}\ }\textbf {\bibinfo {volume} {27}},\ \bibinfo {pages} {485} (\bibinfo {year} {1985})}\BibitemShut {NoStop}%
\bibitem [{\citenamefont {{Rune H{\o}jlund}}(2024)}]{GitHubRepo}%
  \BibitemOpen
  \bibfield  {author} {\bibinfo {author} {\bibnamefont {{Rune H{\o}jlund}}},\ }\href {https://github.com/runehoejlund/plasma-ray-tracer.git} {\enquote {\bibinfo {title} {{GitHub Repository: plasma-ray-tracer}},}\ } (\bibinfo {year} {2024})\BibitemShut {NoStop}%
\bibitem [{\citenamefont {Mazzucato}(2014)}]{Mazzucato2014}%
  \BibitemOpen
  \bibfield  {author} {\bibinfo {author} {\bibfnamefont {E.}~\bibnamefont {Mazzucato}},\ }\href {\doibase 10.1142/9789814571814} {\emph {\bibinfo {title} {Electromagnetic Waves for Thermonuclear Fusion Research}}}\ (\bibinfo  {publisher} {World Scientific Publishing Co.},\ \bibinfo {year} {2014})\ pp.\ \bibinfo {pages} {1--204}\BibitemShut {NoStop}%
\bibitem [{\citenamefont {Dodin}\ \emph {et~al.}(2019)\citenamefont {Dodin}, \citenamefont {Ruiz}, \citenamefont {Yanagihara}, \citenamefont {Zhou},\ and\ \citenamefont {Kubo}}]{Dodin2019}%
  \BibitemOpen
  \bibfield  {author} {\bibinfo {author} {\bibfnamefont {I.~Y.}\ \bibnamefont {Dodin}}, \bibinfo {author} {\bibfnamefont {D.~E.}\ \bibnamefont {Ruiz}}, \bibinfo {author} {\bibfnamefont {K.}~\bibnamefont {Yanagihara}}, \bibinfo {author} {\bibfnamefont {Y.}~\bibnamefont {Zhou}}, \ and\ \bibinfo {author} {\bibfnamefont {S.}~\bibnamefont {Kubo}},\ }\href {\doibase 10.1063/1.5095076} {\bibfield  {journal} {\bibinfo  {journal} {Physics of Plasmas}\ }\textbf {\bibinfo {volume} {26}} (\bibinfo {year} {2019}),\ 10.1063/1.5095076}\BibitemShut {NoStop}%
\bibitem [{\citenamefont {Zhou}\ and\ \citenamefont {Dodin}(2019)}]{Dodin2019Supplementary}%
  \BibitemOpen
  \bibfield  {author} {\bibinfo {author} {\bibfnamefont {Y.}~\bibnamefont {Zhou}}\ and\ \bibinfo {author} {\bibfnamefont {I.~Y.}\ \bibnamefont {Dodin}},\ }\href {\doibase 10.1063/1.5095076} {\enquote {\bibinfo {title} {{Supplementary material: Weyl calculus on a curved configuration space.}}}\ } (\bibinfo {year} {2019})\BibitemShut {NoStop}%
\bibitem [{\citenamefont {Goldstein}\ \emph {et~al.}(2002)\citenamefont {Goldstein}, \citenamefont {Poole},\ and\ \citenamefont {Safko}}]{Goldstein2002}%
  \BibitemOpen
  \bibfield  {author} {\bibinfo {author} {\bibfnamefont {H.}~\bibnamefont {Goldstein}}, \bibinfo {author} {\bibfnamefont {C.}~\bibnamefont {Poole}}, \ and\ \bibinfo {author} {\bibfnamefont {J.}~\bibnamefont {Safko}},\ }\href@noop {} {\emph {\bibinfo {title} {{Classical mechanics}}}},\ \bibinfo {edition} {3rd}\ ed.\ (\bibinfo  {publisher} {Pearson / Addison-Wesley},\ \bibinfo {address} {Upper Saddle River, NJ},\ \bibinfo {year} {2002})\BibitemShut {NoStop}%
\bibitem [{\citenamefont {{Jorge Nocedal}}\ and\ \citenamefont {{Stephen J. Wright}}(2006)}]{Nocedal2006}%
  \BibitemOpen
  \bibfield  {author} {\bibinfo {author} {\bibnamefont {{Jorge Nocedal}}}\ and\ \bibinfo {author} {\bibnamefont {{Stephen J. Wright}}},\ }\href {\doibase 10.1007/978-0-387-40065-5} {\emph {\bibinfo {title} {{Numerical Optimization}}}}\ (\bibinfo  {publisher} {Springer},\ \bibinfo {year} {2006})\BibitemShut {NoStop}%
\bibitem [{\citenamefont {{The SciPy Community}}(2024)}]{SciPySolveIVP}%
  \BibitemOpen
  \bibfield  {author} {\bibinfo {author} {\bibnamefont {{The SciPy Community}}},\ }\href {https://docs.scipy.org/doc/scipy/reference/generated/scipy.integrate.solve_ivp.html} {\enquote {\bibinfo {title} {{SciPy IVP Solver Documentation}},}\ } (\bibinfo {year} {2024})\BibitemShut {NoStop}%
\bibitem [{\citenamefont {Paszke}\ \emph {et~al.}(2019)\citenamefont {Paszke}, \citenamefont {Gross}, \citenamefont {Massa}, \citenamefont {Lerer}, \citenamefont {Bradbury}, \citenamefont {Chanan}, \citenamefont {Killeen}, \citenamefont {Lin}, \citenamefont {Gimelshein}, \citenamefont {Antiga}, \citenamefont {Desmaison}, \citenamefont {K{\"{o}}pf}, \citenamefont {Yang}, \citenamefont {DeVito}, \citenamefont {Raison}, \citenamefont {Tejani}, \citenamefont {Chilamkurthy}, \citenamefont {Steiner}, \citenamefont {Fang}, \citenamefont {Bai},\ and\ \citenamefont {Chintala}}]{PyTorch}%
  \BibitemOpen
  \bibfield  {author} {\bibinfo {author} {\bibfnamefont {A.}~\bibnamefont {Paszke}}, \bibinfo {author} {\bibfnamefont {S.}~\bibnamefont {Gross}}, \bibinfo {author} {\bibfnamefont {F.}~\bibnamefont {Massa}}, \bibinfo {author} {\bibfnamefont {A.}~\bibnamefont {Lerer}}, \bibinfo {author} {\bibfnamefont {J.}~\bibnamefont {Bradbury}}, \bibinfo {author} {\bibfnamefont {G.}~\bibnamefont {Chanan}}, \bibinfo {author} {\bibfnamefont {T.}~\bibnamefont {Killeen}}, \bibinfo {author} {\bibfnamefont {Z.}~\bibnamefont {Lin}}, \bibinfo {author} {\bibfnamefont {N.}~\bibnamefont {Gimelshein}}, \bibinfo {author} {\bibfnamefont {L.}~\bibnamefont {Antiga}}, \bibinfo {author} {\bibfnamefont {A.}~\bibnamefont {Desmaison}}, \bibinfo {author} {\bibfnamefont {A.}~\bibnamefont {K{\"{o}}pf}}, \bibinfo {author} {\bibfnamefont {E.}~\bibnamefont {Yang}}, \bibinfo {author} {\bibfnamefont {Z.}~\bibnamefont {DeVito}}, \bibinfo {author} {\bibfnamefont {M.}~\bibnamefont {Raison}}, \bibinfo {author} {\bibfnamefont {A.}~\bibnamefont {Tejani}}, \bibinfo {author} {\bibfnamefont {S.}~\bibnamefont {Chilamkurthy}}, \bibinfo {author} {\bibfnamefont {B.}~\bibnamefont {Steiner}}, \bibinfo {author} {\bibfnamefont {L.}~\bibnamefont {Fang}}, \bibinfo {author} {\bibfnamefont {J.}~\bibnamefont {Bai}}, \ and\ \bibinfo {author} {\bibfnamefont {S.}~\bibnamefont {Chintala}},\ }\href {\doibase 10.48550/arxiv.1912.01703} {\enquote {\bibinfo {title} {{PyTorch: An Imperative Style, High-Performance Deep Learning Library}},}\ } (\bibinfo {year} {2019})\BibitemShut {NoStop}%
\bibitem [{\citenamefont {Gil}\ \emph {et~al.}(2007)\citenamefont {Gil}, \citenamefont {Segura},\ and\ \citenamefont {Temme}}]{Gil2007}%
  \BibitemOpen
  \bibfield  {author} {\bibinfo {author} {\bibfnamefont {A.}~\bibnamefont {Gil}}, \bibinfo {author} {\bibfnamefont {J.}~\bibnamefont {Segura}}, \ and\ \bibinfo {author} {\bibfnamefont {N.~M.}\ \bibnamefont {Temme}},\ }\href {\doibase 10.1137/1.9780898717822} {\emph {\bibinfo {title} {{Numerical Methods for Special Functions}}}}\ (\bibinfo  {publisher} {Society for Industrial and Applied Mathematics},\ \bibinfo {year} {2007})\BibitemShut {NoStop}%
\bibitem [{\citenamefont {Berrut}\ and\ \citenamefont {Klein}(2014)}]{Berrut2014}%
  \BibitemOpen
  \bibfield  {author} {\bibinfo {author} {\bibfnamefont {J.-P.}\ \bibnamefont {Berrut}}\ and\ \bibinfo {author} {\bibfnamefont {G.}~\bibnamefont {Klein}},\ }\href {\doibase https://doi.org/10.1016/j.cam.2013.03.044} {\bibfield  {journal} {\bibinfo  {journal} {Journal of Computational and Applied Mathematics}\ }\textbf {\bibinfo {volume} {259}},\ \bibinfo {pages} {95} (\bibinfo {year} {2014})},\ \bibinfo {note} {proceedings of the Sixteenth International Congress on Computational and Applied Mathematics (ICCAM-2012), Ghent, Belgium, 9-13 July, 2012}\BibitemShut {NoStop}%
\bibitem [{\citenamefont {Nakatsukasa}\ \emph {et~al.}(2018)\citenamefont {Nakatsukasa}, \citenamefont {Sète},\ and\ \citenamefont {Trefethen}}]{Nakatsukasa2018}%
  \BibitemOpen
  \bibfield  {author} {\bibinfo {author} {\bibfnamefont {Y.}~\bibnamefont {Nakatsukasa}}, \bibinfo {author} {\bibfnamefont {O.}~\bibnamefont {Sète}}, \ and\ \bibinfo {author} {\bibfnamefont {L.~N.}\ \bibnamefont {Trefethen}},\ }\href {\doibase 10.1137/16M1106122} {\bibfield  {journal} {\bibinfo  {journal} {SIAM journal on scientific computing}\ }\textbf {\bibinfo {volume} {40}},\ \bibinfo {pages} {A1494} (\bibinfo {year} {2018})}\BibitemShut {NoStop}%
\bibitem [{\citenamefont {Hofreither}(2021)}]{Hofreither2021}%
  \BibitemOpen
  \bibfield  {author} {\bibinfo {author} {\bibfnamefont {C.}~\bibnamefont {Hofreither}},\ }\href {\doibase 10.1007/s11075-020-01042-0} {\bibfield  {journal} {\bibinfo  {journal} {Numerical algorithms}\ }\textbf {\bibinfo {volume} {88}},\ \bibinfo {pages} {365} (\bibinfo {year} {2021})}\BibitemShut {NoStop}%
\bibitem [{\citenamefont {Trefethen}(2023)}]{Trefethen2023}%
  \BibitemOpen
  \bibfield  {author} {\bibinfo {author} {\bibfnamefont {L.~N.}\ \bibnamefont {Trefethen}},\ }\href {\doibase 10.1007/s13160-023-00599-2} {\bibfield  {journal} {\bibinfo  {journal} {Japan Journal of Industrial and Applied Mathematics}\ } (\bibinfo {year} {2023}),\ 10.1007/s13160-023-00599-2}\BibitemShut {NoStop}%
\bibitem [{\citenamefont {{Hofreither, Clemens}}(2024)}]{baryratGitHub}%
  \BibitemOpen
  \bibfield  {author} {\bibinfo {author} {\bibnamefont {{Hofreither, Clemens}}},\ }\href {https://github.com/c-f-h/baryrat} {\enquote {\bibinfo {title} {{GitHub Repository: baryrat}},}\ } (\bibinfo {year} {2024})\BibitemShut {NoStop}%
\bibitem [{\citenamefont {Griffiths}(2018)}]{Griffiths2018}%
  \BibitemOpen
  \bibfield  {author} {\bibinfo {author} {\bibfnamefont {D.~J.}\ \bibnamefont {Griffiths}},\ }\href@noop {} {\emph {\bibinfo {title} {{Introduction to quantum mechanics}}}},\ \bibinfo {edition} {third edition.}\ ed.\ (\bibinfo  {publisher} {Cambridge University Press},\ \bibinfo {address} {Cambridge, United Kingdom},\ \bibinfo {year} {2018})\BibitemShut {NoStop}%
\bibitem [{\citenamefont {Laqua}\ \emph {et~al.}(1997)\citenamefont {Laqua}, \citenamefont {Erckmann}, \citenamefont {Hartfu\ss{}},\ and\ \citenamefont {Laqua}}]{Laqua1997}%
  \BibitemOpen
  \bibfield  {author} {\bibinfo {author} {\bibfnamefont {H.~P.}\ \bibnamefont {Laqua}}, \bibinfo {author} {\bibfnamefont {V.}~\bibnamefont {Erckmann}}, \bibinfo {author} {\bibfnamefont {H.~J.}\ \bibnamefont {Hartfu\ss{}}}, \ and\ \bibinfo {author} {\bibfnamefont {H.}~\bibnamefont {Laqua}} (\bibinfo {collaboration} {W7-AS Team ECRH Group}),\ }\href {\doibase 10.1103/PhysRevLett.78.3467} {\bibfield  {journal} {\bibinfo  {journal} {Phys. Rev. Lett.}\ }\textbf {\bibinfo {volume} {78}},\ \bibinfo {pages} {3467} (\bibinfo {year} {1997})}\BibitemShut {NoStop}%
\bibitem [{\citenamefont {Guo}\ \emph {et~al.}(2022)\citenamefont {Guo}, \citenamefont {Ashida}, \citenamefont {Noguchi}, \citenamefont {Kajita}, \citenamefont {Tanaka}, \citenamefont {Uchida},\ and\ \citenamefont {Maekawa}}]{Guo2022}%
  \BibitemOpen
  \bibfield  {author} {\bibinfo {author} {\bibfnamefont {X.}~\bibnamefont {Guo}}, \bibinfo {author} {\bibfnamefont {R.}~\bibnamefont {Ashida}}, \bibinfo {author} {\bibfnamefont {Y.}~\bibnamefont {Noguchi}}, \bibinfo {author} {\bibfnamefont {R.}~\bibnamefont {Kajita}}, \bibinfo {author} {\bibfnamefont {H.}~\bibnamefont {Tanaka}}, \bibinfo {author} {\bibfnamefont {M.}~\bibnamefont {Uchida}}, \ and\ \bibinfo {author} {\bibfnamefont {T.}~\bibnamefont {Maekawa}},\ }\href {\doibase 10.1088/1361-6587/ac4522} {\bibfield  {journal} {\bibinfo  {journal} {Plasma Physics and Controlled Fusion}\ }\textbf {\bibinfo {volume} {64}},\ \bibinfo {pages} {035008} (\bibinfo {year} {2022})}\BibitemShut {NoStop}%
\bibitem [{\citenamefont {Clod}\ \emph {et~al.}(2024)\citenamefont {Clod}, \citenamefont {Senstius}, \citenamefont {Nielsen}, \citenamefont {Ragona}, \citenamefont {Thrys\o{}e}, \citenamefont {Kumar}, \citenamefont {Coda},\ and\ \citenamefont {Nielsen}}]{Clod2024}%
  \BibitemOpen
  \bibfield  {author} {\bibinfo {author} {\bibfnamefont {A.}~\bibnamefont {Clod}}, \bibinfo {author} {\bibfnamefont {M.~G.}\ \bibnamefont {Senstius}}, \bibinfo {author} {\bibfnamefont {A.~H.}\ \bibnamefont {Nielsen}}, \bibinfo {author} {\bibfnamefont {R.}~\bibnamefont {Ragona}}, \bibinfo {author} {\bibfnamefont {A.~S.}\ \bibnamefont {Thrys\o{}e}}, \bibinfo {author} {\bibfnamefont {U.}~\bibnamefont {Kumar}}, \bibinfo {author} {\bibfnamefont {S.}~\bibnamefont {Coda}}, \ and\ \bibinfo {author} {\bibfnamefont {S.~K.}\ \bibnamefont {Nielsen}} (\bibinfo {collaboration} {The TCV team}),\ }\href {\doibase 10.1103/PhysRevLett.132.135101} {\bibfield  {journal} {\bibinfo  {journal} {Phys. Rev. Lett.}\ }\textbf {\bibinfo {volume} {132}},\ \bibinfo {pages} {135101} (\bibinfo {year} {2024})}\BibitemShut {NoStop}%
\bibitem [{\citenamefont {Kudlacek}\ \emph {et~al.}(2024)\citenamefont {Kudlacek}, \citenamefont {David}, \citenamefont {Gomez}, \citenamefont {Gräter}, \citenamefont {Sieglin}, \citenamefont {Treutterer}, \citenamefont {Weiland}, \citenamefont {Zehetbauer}, \citenamefont {Berkel}, \citenamefont {Bernert}, \citenamefont {Bosman}, \citenamefont {Felici}, \citenamefont {Giannone}, \citenamefont {Illerhaus}, \citenamefont {Kropackova}, \citenamefont {Lang}, \citenamefont {Maraschek}, \citenamefont {Ploeckl}, \citenamefont {Reich},\ and\ \citenamefont {Kubincova}}]{Kudlacek2024}%
  \BibitemOpen
  \bibfield  {author} {\bibinfo {author} {\bibfnamefont {O.}~\bibnamefont {Kudlacek}}, \bibinfo {author} {\bibfnamefont {P.}~\bibnamefont {David}}, \bibinfo {author} {\bibfnamefont {I.}~\bibnamefont {Gomez}}, \bibinfo {author} {\bibfnamefont {A.}~\bibnamefont {Gräter}}, \bibinfo {author} {\bibfnamefont {B.}~\bibnamefont {Sieglin}}, \bibinfo {author} {\bibfnamefont {W.}~\bibnamefont {Treutterer}}, \bibinfo {author} {\bibfnamefont {M.}~\bibnamefont {Weiland}}, \bibinfo {author} {\bibfnamefont {T.}~\bibnamefont {Zehetbauer}}, \bibinfo {author} {\bibfnamefont {M.~V.}\ \bibnamefont {Berkel}}, \bibinfo {author} {\bibfnamefont {M.}~\bibnamefont {Bernert}}, \bibinfo {author} {\bibfnamefont {T.}~\bibnamefont {Bosman}}, \bibinfo {author} {\bibfnamefont {F.}~\bibnamefont {Felici}}, \bibinfo {author} {\bibfnamefont {L.}~\bibnamefont {Giannone}}, \bibinfo {author} {\bibfnamefont {J.}~\bibnamefont {Illerhaus}}, \bibinfo {author} {\bibfnamefont {D.}~\bibnamefont {Kropackova}}, \bibinfo {author} {\bibfnamefont {P.}~\bibnamefont {Lang}}, \bibinfo {author} {\bibfnamefont {M.}~\bibnamefont {Maraschek}}, \bibinfo {author} {\bibfnamefont {B.}~\bibnamefont {Ploeckl}}, \bibinfo {author} {\bibfnamefont {M.}~\bibnamefont {Reich}}, \ and\ \bibinfo {author} {\bibfnamefont {A.~V.}\ \bibnamefont {Kubincova}},\ }\href {\doibase 10.1088/1741-4326/ad3472} {\bibfield  {journal} {\bibinfo  {journal} {Nuclear Fusion}\ }\textbf {\bibinfo {volume} {64}},\ \bibinfo {pages} {056012} (\bibinfo {year} {2024})}\BibitemShut {NoStop}%
\bibitem [{\citenamefont {Hansen}\ \emph {et~al.}(2019)\citenamefont {Hansen}, \citenamefont {Nielsen}, \citenamefont {Stober}, \citenamefont {Rasmussen}, \citenamefont {Salewski},\ and\ \citenamefont {Stejner}}]{Hansen2019b}%
  \BibitemOpen
  \bibfield  {author} {\bibinfo {author} {\bibfnamefont {S.~K.}\ \bibnamefont {Hansen}}, \bibinfo {author} {\bibfnamefont {S.~K.}\ \bibnamefont {Nielsen}}, \bibinfo {author} {\bibfnamefont {J.}~\bibnamefont {Stober}}, \bibinfo {author} {\bibfnamefont {J.}~\bibnamefont {Rasmussen}}, \bibinfo {author} {\bibfnamefont {M.}~\bibnamefont {Salewski}}, \ and\ \bibinfo {author} {\bibfnamefont {M.}~\bibnamefont {Stejner}},\ }\href {\doibase 10.1063/1.5091659} {\bibfield  {journal} {\bibinfo  {journal} {Physics of Plasmas}\ }\textbf {\bibinfo {volume} {26}} (\bibinfo {year} {2019}),\ 10.1063/1.5091659}\BibitemShut {NoStop}%
\bibitem [{\citenamefont {Arber}\ \emph {et~al.}(2015)\citenamefont {Arber}, \citenamefont {Bennett}, \citenamefont {Brady}, \citenamefont {Lawrence-Douglas}, \citenamefont {Ramsay}, \citenamefont {Sircombe}, \citenamefont {Gillies}, \citenamefont {Evans}, \citenamefont {Schmitz}, \citenamefont {Bell},\ and\ \citenamefont {Ridgers}}]{Arber2015}%
  \BibitemOpen
  \bibfield  {author} {\bibinfo {author} {\bibfnamefont {T.~D.}\ \bibnamefont {Arber}}, \bibinfo {author} {\bibfnamefont {K.}~\bibnamefont {Bennett}}, \bibinfo {author} {\bibfnamefont {C.~S.}\ \bibnamefont {Brady}}, \bibinfo {author} {\bibfnamefont {A.}~\bibnamefont {Lawrence-Douglas}}, \bibinfo {author} {\bibfnamefont {M.~G.}\ \bibnamefont {Ramsay}}, \bibinfo {author} {\bibfnamefont {N.~J.}\ \bibnamefont {Sircombe}}, \bibinfo {author} {\bibfnamefont {P.}~\bibnamefont {Gillies}}, \bibinfo {author} {\bibfnamefont {R.~G.}\ \bibnamefont {Evans}}, \bibinfo {author} {\bibfnamefont {H.}~\bibnamefont {Schmitz}}, \bibinfo {author} {\bibfnamefont {A.~R.}\ \bibnamefont {Bell}}, \ and\ \bibinfo {author} {\bibfnamefont {C.~P.}\ \bibnamefont {Ridgers}},\ }\href {\doibase 10.1088/0741-3335/57/11/113001} {\bibfield  {journal} {\bibinfo  {journal} {Plasma Physics and Controlled Fusion}\ }\textbf {\bibinfo {volume} {57}},\ \bibinfo {pages} {113001} (\bibinfo {year} {2015})}\BibitemShut {NoStop}%
\bibitem [{\citenamefont {Senstius}\ \emph {et~al.}(2020)\citenamefont {Senstius}, \citenamefont {Nielsen}, \citenamefont {Vann},\ and\ \citenamefont {Hansen}}]{Senstius2020b}%
  \BibitemOpen
  \bibfield  {author} {\bibinfo {author} {\bibfnamefont {M.~G.}\ \bibnamefont {Senstius}}, \bibinfo {author} {\bibfnamefont {S.~K.}\ \bibnamefont {Nielsen}}, \bibinfo {author} {\bibfnamefont {R.~G.}\ \bibnamefont {Vann}}, \ and\ \bibinfo {author} {\bibfnamefont {S.~K.}\ \bibnamefont {Hansen}},\ }\href {\doibase 10.1088/1361-6587/ab49ca} {\bibfield  {journal} {\bibinfo  {journal} {Plasma Physics and Controlled Fusion}\ }\textbf {\bibinfo {volume} {62}},\ \bibinfo {pages} {025010} (\bibinfo {year} {2020})}\BibitemShut {NoStop}%
\bibitem [{\citenamefont {Senstius}\ \emph {et~al.}(2024)\citenamefont {Senstius}, \citenamefont {Freethy},\ and\ \citenamefont {Nielsen}}]{Senstius2024}%
  \BibitemOpen
  \bibfield  {author} {\bibinfo {author} {\bibfnamefont {M.~G.}\ \bibnamefont {Senstius}}, \bibinfo {author} {\bibfnamefont {S.~J.}\ \bibnamefont {Freethy}}, \ and\ \bibinfo {author} {\bibfnamefont {S.~K.}\ \bibnamefont {Nielsen}},\ }\href {\doibase 10.1063/5.0187071} {\bibfield  {journal} {\bibinfo  {journal} {Physics of Plasmas}\ }\textbf {\bibinfo {volume} {31}},\ \bibinfo {pages} {032308} (\bibinfo {year} {2024})}\BibitemShut {NoStop}%
\bibitem [{\citenamefont {Crawford}\ and\ \citenamefont {Tataronis}(1965)}]{Crawford1965}%
  \BibitemOpen
  \bibfield  {author} {\bibinfo {author} {\bibfnamefont {F.~W.}\ \bibnamefont {Crawford}}\ and\ \bibinfo {author} {\bibfnamefont {J.~A.}\ \bibnamefont {Tataronis}},\ }\href {\doibase 10.1063/1.1714609} {\bibfield  {journal} {\bibinfo  {journal} {Journal of Applied Physics}\ }\textbf {\bibinfo {volume} {36}},\ \bibinfo {pages} {2930} (\bibinfo {year} {1965})}\BibitemShut {NoStop}%
\bibitem [{\citenamefont {Hansen}(2019)}]{Hansen2019}%
  \BibitemOpen
  \bibfield  {author} {\bibinfo {author} {\bibfnamefont {S.~K.}\ \bibnamefont {Hansen}},\ }\emph {\bibinfo {title} {Parametric Decay Instabilities in the Electron Cyclotron Resonance Heating Beams at ASDEX Upgrade}},\ \href {https://orbit.dtu.dk/en/publications/parametric-decay-instabilities-in-the-electron-cyclotron-resonanc} {Ph.D. thesis},\ \bibinfo  {school} {Technical University of Denmark} (\bibinfo {year} {2019})\BibitemShut {NoStop}%
\bibitem [{\citenamefont {Senstius}\ \emph {et~al.}(2021)\citenamefont {Senstius}, \citenamefont {Nielsen},\ and\ \citenamefont {Vann}}]{Senstius2021}%
  \BibitemOpen
  \bibfield  {author} {\bibinfo {author} {\bibfnamefont {M.~G.}\ \bibnamefont {Senstius}}, \bibinfo {author} {\bibfnamefont {S.~K.}\ \bibnamefont {Nielsen}}, \ and\ \bibinfo {author} {\bibfnamefont {R.~G.~L.}\ \bibnamefont {Vann}},\ }\href {\doibase 10.1088/1361-6587/abf85a} {\bibfield  {journal} {\bibinfo  {journal} {Plasma Physics and Controlled Fusion}\ }\textbf {\bibinfo {volume} {63}},\ \bibinfo {pages} {065018} (\bibinfo {year} {2021})}\BibitemShut {NoStop}%
\end{thebibliography}%

\newpage
\end{document}